\documentclass[aps,reprint,english,showpacs,pra,groupedaddress,floatfix]{revtex4-1}
\usepackage[T1]{fontenc}
\usepackage[utf8]{inputenc}
\usepackage{babel}
\usepackage{color}
\usepackage{amsmath}
\usepackage{amssymb}
\usepackage{graphicx}
\usepackage{subscript}

\begin{document}

\title{Transverse distinguishability of entangled photons with arbitrarily
shaped spatial near- and far-field distributions}

\author{Robert Elsner }

\email{robert.elsner@uni-potsdam.de}

\affiliation{University of Potsdam, Institute for Physics and Astronomy, Karl-Liebknecht-Str.
24/25, 14476 Potsdam, Germany}

\author{Dirk Puhlmann}
\affiliation{University of Potsdam, Institute for Physics and Astronomy, Karl-Liebknecht-Str.
24/25, 14476 Potsdam, Germany}

\author{Gregor Pieplow}
\affiliation{University of Potsdam, Institute for Physics and Astronomy, Karl-Liebknecht-Str.
24/25, 14476 Potsdam, Germany}

\author{Axel Heuer}
\affiliation{University of Potsdam, Institute for Physics and Astronomy, Karl-Liebknecht-Str.
24/25, 14476 Potsdam, Germany}

\author{Ralf Menzel}

\email{photonics@uni-potsdam.de}

\affiliation{University of Potsdam, Institute for Physics and Astronomy, Karl-Liebknecht-Str.
24/25, 14476 Potsdam, Germany}

\date{\today}
\begin{abstract}
Entangled photons generated by spontaneous parametric down conversion
inside a nonlinear crystal exhibit a complex spatial photon count
distribution. A quantitative description of this distribution helps
with the interpretation of experiments that depend on this structure.
We developed a theoretical model and an accompanying numerical calculation
that includes the effects of phase matching and the crystal properties
to describe a wide range of spatial effects in two-photon experiments.
The numerical calculation was tested against selected analytical approximations.
We furthermore performed a double-slit experiment where we measured
the visibility $V$ and the distinguishability $D$ and obtained $D^{2}+V^{2}=1.43$.
The numerical model accurately predicts these experimental results.
\end{abstract}

\pacs{42.50.Tx, 42.50.Xa, 42.50.Ar, 03.65.Ud, 02.70.-c}

\maketitle

\section{Introduction}

\begin{figure}
\begin{centering}
\includegraphics[width=0.8\columnwidth]{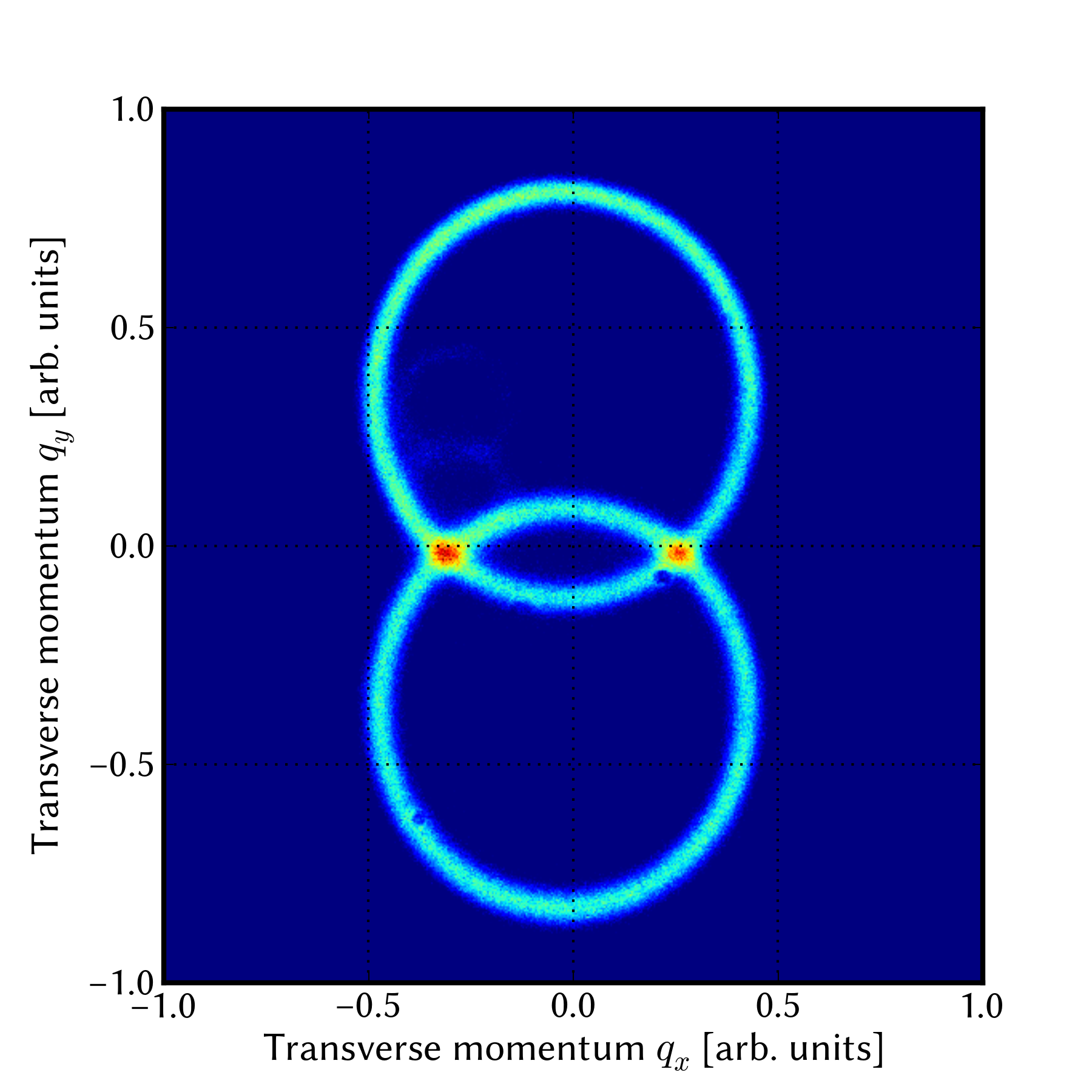}
\par\end{centering}

\caption{Typical far-field cones of the signal and idler photons generated
by SPDC as measured in a plane orthogonal to the propagation direction
of the pump beam.\label{fig:Typical-far-field-cone}}

\end{figure}

A common source of entangled photon pairs is spontaneous parametric
down conversion (SPDC)\citep{burnham_observation_1970}, where a pump
photon with frequency $\omega_{p}$, incident on a nonlinear crystal
is split into two new photons. The two photons, usually referred to
as signal and idler are at lower frequencies $\omega_{s},\omega_{i}<\omega_{p}$
and are entangled in multiple degrees of freedom such as energy, angular
momentum and polarization. Depending on the alignment of the pump
photon wave vector with respect to the optical axis of the crystal,
typical cone structures of the emission directions of signal and idler
photons can be observed (Fig. \ref{fig:Typical-far-field-cone}) in
the Fraunhofer far-field. These are the geometric manifestation of
the phase matching condition in combination with energy and momentum
conservation of the signal, idler and pump photons, that maximize
the efficiency of the down conversion. Due to the spatial separation
of the entangled photons in the cones, SPDC is a useful tool to investigate
non-classical states of light \citep{ou_observation_1988,fonseca_measurement_1999,friberg_measurement_1985,ghosh_observation_1987,hong_experimental_1986},
to explore possible applications in quantum information \citep{neves_generation_2005}
or to probe the foundations of quantum mechanics such as in the Einstein-Podolsky-Rosen
paradox \citep{einstein_can_1935,howell_realization_2004}.

Comprehensive theoretical descriptions of the SPDC process were developed
by Mollow \citep{mollow_photon_1973} and Hong and Mandel \citep{hong_theory_1985}.
An in-depth review was written by Walborn et al. \citep{walborn_spatial_2010}.
These works serve as starting point for our model, that will make
them applicable to describe our experiments. Our theoretical considerations
allow for apertures in the near- and far-field of the nonlinear medium
to manipulate the angular momentum distribution of the SPDC light.
Such a framework supports the design and interpretation of experiments
that study the transverse spatial structure of SPDC light in two dimensions
in such detail for the first time. Related work albeit partially without
considering the phase matching condition and restricted to one spatial
dimension instead of two as in our case was published by Abouraddy
et al. \citep{abouraddy_entangled-photon_2002}. The far-field structure
of SPDC light in two dimensions has been investigated by Bennink et
al. \citep{bennink_spatial_2006} but without our more involved inclusion
of the near-field plane.

In this paper we specifically focus on spatial coincidence measurements
of SPDC light. In the first section \ref{sub:Theoretical-Model} of
this paper we present a theory that encompasses apertures in far and
near field of the SPDC light and provides an expression for said coincidences.
Their evaluation necessitates solving quite involved integral expressions.
We therefore implemented a numerical simulation to obtain quantitative
approximations that can be compared to experimental data. In section
\ref{sub:Selected-Analytic-Solutions} a set of analytical solutions
that follow from various approximations are derived. These analytical
solutions are used as test cases for the numerical simulation. We
conclude this section with a description of a double-slit experiment
with entangled photons performed by Menzel et al.\citep{menzel_wave-particle_2012,menzel_two-photon_2013}.
We repeat this experiment which serves as a first comparison with
empirical data. We finally establish in section \ref{sub:Experimental-results-and}
a limit for an inequality linking the visibility $V$ of the interference
fringes and the which-way distinguishability $D$ at the double slit.
We observe that our experiment is closer to this limit as previously
reported in \citep{menzel_wave-particle_2012,menzel_two-photon_2013}.
This new numerical model will enable a detailed two-dimensional investigation
of the fair sampling problem at a double slit \citep{leach_duality_2014}
for both type I and type II phase matching.

\section{Materials and Methods}

\subsection{Theoretical Model\label{sub:Theoretical-Model}}

\begin{figure}
\begin{centering}
\includegraphics[width=1\columnwidth]{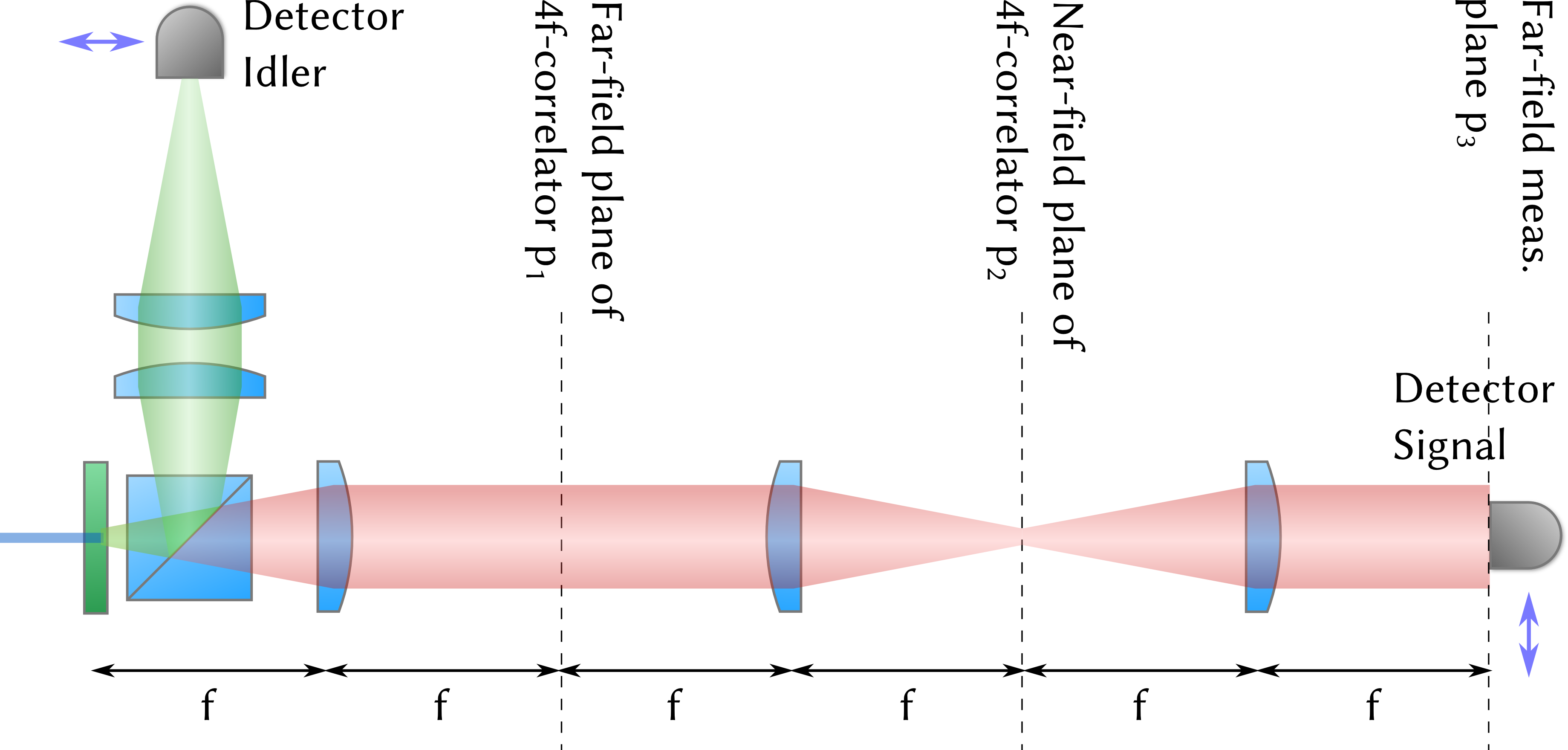}
\par\end{centering}

\caption{Schematic of a possible realization of an experiment described by
our model. The entangled photons from a type II SPDC source are separated
by a polarizing beam splitter (PBS) and manipulated separately by
4f-correlators. The idler path (green) is identical to the signal
path (red) but is not shown to scale.\label{fig:schematic_model}}
\end{figure}

The theoretical considerations presented in this section follow the
ideas outlined in \citep{walborn_spatial_2010}, which we\textcolor{red}{{}
}adapted to accommodate the numerical analysis of spatial coincidence
events. We start with describing the quantum state that arises from
the SPDC process and we review the principal approximations. We then
introduce expressions for the quantized electric field operators and
the two-photon coincidence count rates which include the effects of
optical elements between the crystal and the detectors. We use the
terms ``coincidence count rate'' or ``two-photon count rate''
to describe the - unnormalized - probability of simultaneously detecting
a signal and idler photon at given positions of the corresponding
detectors.

The scope of the model is to encompass entangled two-photon experiments,
where coincidence count rates are measured. We further narrow our
model to a setup where two independent 4f-correlators are placed into
the signal and idler beam of an SPDC source as can be seen in Fig.
\ref{fig:schematic_model}. In the respective focal planes the 4f-correlators
facilitate a Fourier-transformation of the incident field \citep{stoler_operator_1982}.
The 4f-correlator allows us to access the near-field and the angular
transverse spectrum in a single setup. The Detectors can then be positioned
in the near- or far-field behind the 4f-correlators. They measure
the coincidence counts in the planes $p_{1}^{(s/i)},p_{2}^{(s/i)}$
orthogonal to the propagation direction of the respective beam. The
superscripts indicate if the plane is positioned in the idler or signal
path. We omit the superscript in the remainder of this paper if an
argument applies to both the signal and idler paths. For further reference
see Fig. \ref{fig:schematic_model}. 

The signal and idler beams are separated with a polarizing beam splitter
(PBS). Narrow band spectral filters are placed into both paths such
that $\omega_{p}=\omega_{s}+\omega_{i}$. The angular momentum spectrum
of the idler and signal beam can be manipulated independently in the
far-field plane $p_{1}$ of the 4f-correlator by inserting arbitrary
amplitude masks $\mathcal{T}_{s/i}(\boldsymbol{q}_{s/i}):\mathbb{R}^{2}\rightarrow\mathbb{R}$.
Here $\boldsymbol{q_{s/i}}$ denotes the wavevector component of the
signal/idler beam, transverse to the propagation direction. Because
the first lens of the 4f-correlator establishes a simple linear mapping
$\boldsymbol{q}(\boldsymbol{\rho}_{p_{1}})=\xi\boldsymbol{\rho}_{p_{1}}$
between the spatial ($\boldsymbol{\rho}_{p_{1}}$) and momentum ($\boldsymbol{q}$)
coordinates in the far-field plane $p_{1}$ of the 4f-correlator,
we pretend that we are manipulating the angular spectrum directly.
The crystal near field is then imaged onto the near field plane $p_{2}$
of the 4f-correlator with a second amplitude mask $\mathcal{N}_{s/i}(\boldsymbol{\rho}_{s/i}):\mathbb{R}^{2}\rightarrow\mathbb{R}$.
The signal and idler detectors can be placed directly behind the near-field
amplitude mask at $p_{2}$ or in the far field measurement plane $p_{3}$.
Arbitrary combinations of positions of the two detectors are possible.
The coordinates at $p_{1}$ and $p_{3}$ are denoted by $\boldsymbol{q}_{s/i}$
and $\boldsymbol{q}_{s/i}'$ respectively while the spatial coordinate
at $p_{2}$ is $\boldsymbol{\rho}_{s/i}$. The transverse momentum
coordinate of the pump beam at the plane through the center of the
nonlinear crystal is $\boldsymbol{q}$ without a subscript with $\boldsymbol{\rho}$
the associated spatial coordinate.

With these conventions the quantum state of the entangled photons
created by SPDC is \citep{walborn_spatial_2010} 
\begin{equation}
\left|\psi\right\rangle =G\iint d\boldsymbol{q}_{s}d\boldsymbol{q}_{i}\tilde{\Phi}(\boldsymbol{q}_{s},\boldsymbol{q}_{i})\left|\boldsymbol{q}_{s}\right\rangle \left|\boldsymbol{q}_{i}\right\rangle \label{eq:state}
\end{equation}
The state $|\boldsymbol{q}_{s/i}\rangle$ represents a single-photon
Fock state, that is characterized by its transverse momentum $\boldsymbol{q}_{s/i}$
while $k_{z}$ is fixed by the dispersion relation $k^{2}=\omega$.
The constant factor G depends on the efficiency of the conversion process
among other parameters which are of little interest in the context
of this paper. Henceforth are going to omit these proportionality
constants. The amplitude for the two photon state $\tilde{\Phi}(\boldsymbol{q}_{s},\boldsymbol{q}_{i})$
is given by
\begin{equation}
\tilde{\Phi}(\boldsymbol{q}_{s},\boldsymbol{q}_{i})=\tilde{u}(\boldsymbol{q}_{s}+\boldsymbol{q}_{i})\text{sinc}(\Delta k_{z}L/2)\label{eq:spectral_amplitude}
\end{equation}

where $L$ is the length of the crystal, $\Delta k_{z}=k_{zp}-k_{zs}-k_{zi}$
is the longitudinal wave vector mismatch and $\tilde{u}(\boldsymbol{q})$
is the transverse momentum spectrum of the pump beam. We define the
spectrum according to
\begin{equation}
\tilde{u}(\boldsymbol{q})=\int d\boldsymbol{\rho}u(\boldsymbol{\rho})e^{-i\boldsymbol{\rho}\cdot\boldsymbol{q}}\label{eq:pump_spectrum}
\end{equation}
We can interpret $u(\boldsymbol{\rho})$ as the angular electric field
amplitude distribution of the pump beam. For further details on the
explicit derivation of Eqs. (\ref{eq:state})-(\ref{eq:spectral_amplitude})
one can consult \citep{hong_theory_1985,walborn_spatial_2010}. The
most important assumption that were made in \citep{walborn_spatial_2010}
are:
\begin{enumerate}
\item The state $\left|\psi\right\rangle $ is a first-order approximation
e.g. each constituent photon state $\left|\boldsymbol{q}\right\rangle $
is a single-photon state. Effects that arise from multi-photon generation
are ignored. This assumption is justified if the probability of creating
more than one photon pair during a detection interval is negligible
which is easily realized in SPDC experiments.
\item We assume the pump, signal and idler beams to be monochromatic which
can be approximated experimentally by using narrow-band filters.
\item The quantization volume is large enough to allow the approximation
of sums over $\boldsymbol{k}$ by integrals. The transverse dimensions
of the crystal are larger than the transverse extent of the pump beam.
\item The beams are taken to be linearly polarized with the pump beam having
extraordinary (\textit{e}) polarization and the signal and idler beams
having ordinary (o) polarization for type I phase matching e.g. $e\rightarrow oo$.
For type II phase matching one of the down-converted fields is polarized
along the e-direction e.g. $e\rightarrow oe$.
\item We ignore effects due to diffraction and refraction at the crystal's
surface and instead assume that the crystal is embedded in a medium
with a matched refractive index and no birefringence. Birefringence
inside the nonlinear crystal is taken into account.
\end{enumerate}
The electric field operator in the paraxial approximation for a measurement
directly behind the preparation stage in the plane $p_{2}$ is given
by

\begin{equation}
\hat{\boldsymbol{E}}^{+}(\boldsymbol{\rho})=\mathcal{N}(\boldsymbol{\rho})\int d\boldsymbol{q}\mathcal{T}(\boldsymbol{q})\hat{\boldsymbol{a}}(\boldsymbol{q})\exp(i\boldsymbol{q}\cdot\boldsymbol{\rho})\label{eq:near_field}
\end{equation}

The field operator in the far field is:

\begin{align}
\hat{\boldsymbol{E}}^{+}(\boldsymbol{q}')= & \int d\boldsymbol{\rho}\exp(-i\boldsymbol{q}'\cdot\boldsymbol{\rho})\mathcal{N}(\boldsymbol{\rho})\times\label{eq:op_far_field}\\
\times & \int d\boldsymbol{q}\mathcal{T}(\boldsymbol{q})\hat{\boldsymbol{a}}(\boldsymbol{q})\exp(i\boldsymbol{q}\cdot\boldsymbol{\rho})\nonumber 
\end{align}
 The coincidence detection rate or probability for simultaneously
detecting an idler and a signal photon 
\begin{equation}
C^{(2)}=\left|\hat{\boldsymbol{E}}_{s}^{(+)}\hat{\boldsymbol{E}}_{i}^{(+)}|\psi\rangle\right|^{2}\label{eq:coincidence_rate}
\end{equation}
 Thus inserting (\ref{eq:near_field}) and (\ref{eq:op_far_field})
into (\ref{eq:coincidence_rate}) yields the two-photon coincidence
count rate in the near-field case 
\begin{align}
C^{(2)}(\boldsymbol{\rho}_{s},\boldsymbol{\rho}_{i})= & \left|\mathcal{N}_{s}(\boldsymbol{\rho}_{s})\mathcal{N}_{i}(\boldsymbol{\rho}_{i})\times\right.\label{eq:c2_near_field}\\
\times & \left.\mathcal{F}^{-1}\left\lbrace \tilde{\Phi}(\boldsymbol{q}_{s},\boldsymbol{q}_{i})\mathcal{T}_{s}(\boldsymbol{q}_{s})\mathcal{T}_{i}(\boldsymbol{q}_{i})\right\rbrace \right|^{2}\nonumber 
\end{align}
In the far-field case we find the coincidence rate 
\begin{align}
C^{(2)}(\boldsymbol{q}_{s}',\boldsymbol{q}_{i}')= & \mathcal{F}\left\lbrace \mathcal{N}_{s}(\boldsymbol{\rho}_{s})\mathcal{N}_{i}(\boldsymbol{\rho}_{i})\right.\times\label{eq:c2_far_field}\\
\times & \left.\left.\mathcal{F}^{-1}\left\lbrace \tilde{\Phi}(\boldsymbol{q}_{s},\boldsymbol{q}_{i})\mathcal{T}_{s}(\boldsymbol{q}_{s})\mathcal{T}_{i}(\boldsymbol{q}_{i})\right\rbrace \right\rbrace \right|^{2}\nonumber 
\end{align}
where $\mathcal{F}$ and $\mathcal{F}^{-1}$ denote the forward and
inverse Fourier transform. These expressions are analogous to the
propagation of a classical electric field in the paraxial approximation
with the distinction, that the field amplitude $\tilde{\Phi}$ is
a non-separable function of the transverse coordinates of the signal
and the idler photons. The conditional probability of detecting a
signal or idler photon is obtained by tracing out the idler or signal
coordinate.
\begin{equation}
C^{(1)}(\boldsymbol{\rho}_{s/i})=\int d\boldsymbol{\rho}_{i/s}C^{(2)}(\boldsymbol{\rho}_{s},\boldsymbol{\rho}_{i})
\end{equation}
This corresponds to the presence of a bucket detector in either the
signal or idler paths. If there are no apertures $\mathcal{T}$, $\mathcal{N}$
present in the experimental setup, this quantity can be interpreted
as the single-photon count rate of the signal or idler. In this particular
case we call $C^{(1)}$ the ``single photon count rate''. Otherwise
we refer to it as the reduced coincidence count rate.

Finally, we have to account for the behavior of the longitudinal phase
mismatch $\Delta k_{z}$ which is key for accurately modeling the
spatial behavior of SPDC light. In a birefringent crystal the z-component
of the wave vector of an extraordinary beam can be expressed \citep{walborn_spatial_2010}
in terms of its transverse components as

\begin{equation}
k_{z}\approx\alpha q_{x}+\eta\frac{\omega}{c_{0}}-\frac{c_{0}}{2\eta\omega}(\beta^{2}q_{x}^{2}+\gamma^{2}q_{y}^{2})\label{eq:kz_extraordinary}
\end{equation}
while for an ordinary beam
\begin{equation}
k_{z}\approx n_{o}\frac{\omega}{c_{0}}-\frac{c_{0}}{2n_{o}\omega}|\boldsymbol{q}|^{2}\label{eq:kz_ordinary}
\end{equation}

Usually the quantities $\beta$ and $\gamma$ are close to unity.
They cause a slight astigmatism of the down-converted fields. $\eta$
is the mean refractive index for a beam of angular frequency $\omega$.
The constant $\alpha$ describes the transverse walk-off of the beam
and $n_{0}=n_{o}(\omega)$ is the ordinary refractive index calculated
at the angular frequency $\omega$. These expressions again are derived
using the paraxial approximation.

Thus experiments that can be cast into the framework outlined above
can be modeled by inserting the appropriate field operators (\ref{eq:near_field}),(\ref{eq:op_far_field})
into (\ref{eq:coincidence_rate}) with the quantum state given by
(\ref{eq:state}) and the corresponding longitudinal wave vector components
from Eqs. (\ref{eq:kz_extraordinary}) and (\ref{eq:kz_ordinary}).

\subsection{Testcase Analytic Solutions\label{sub:Selected-Analytic-Solutions}}

The numerical simulation is validated against two analytical test
cases. The first one is the so-called thin crystal approximation which
assumes an infinitely thin crystal with $L=0$. This approximation
removes the phase-matching condition $\Delta k_{z}\approx0$ and results
in an uniform transverse momentum distribution of the SPDC light.
We assume that there are no apertures $\mathcal{T},\mathcal{N}$ inserted
into the path of the SPDC light, the detectors are both placed in
the near-field of the crystal at $p_{2}^{(s/i)}$ and the process
is degenerate with $\omega_{i}=\omega_{s}$. The near-field coincidence
distribution is then derived by inserting (\ref{eq:spectral_amplitude})
into (\ref{eq:c2_near_field})

\begin{equation}
C^{(2)}(\boldsymbol{\rho}_{s},\boldsymbol{\rho}_{i})=\begin{cases}
P|u(\boldsymbol{\rho}_{s})|^{2} & \boldsymbol{\rho}_{s}=\boldsymbol{\rho}_{i}\\
0 & \boldsymbol{\rho}_{s}\neq\boldsymbol{\rho}_{i}
\end{cases}\label{eq:two_photon_thin_crystal_near_field}
\end{equation}
\begin{equation}
C^{(1)}(\boldsymbol{\rho}_{s})=\int d\boldsymbol{\rho}_{i}C^{(2)}(\boldsymbol{\rho}_{s},\boldsymbol{\rho}_{i})=|u(\boldsymbol{\rho}_{s})|^{2}\label{eq:single_photon_thin_crystal_near_field}
\end{equation}

P denotes a constant that depends on the transverse cutoff aperture.
Throughout this section we use P to remind of the nonessential dependence
of the absolute value of the respective expressions on the specific
aperture. The signal and idler photons exhibit perfect correlations
as evident from Eq. (\ref{eq:two_photon_thin_crystal_near_field}).
The single-photon rate as seen from the signal detector reproduces
the transverse intensity profile of the pump beam (\ref{eq:single_photon_thin_crystal_near_field}).
With both detectors in the far-field plane $p_{3}$, the coincidence
count and single photon rates are

\begin{equation}
C^{(2)}(\boldsymbol{q}_{s},\boldsymbol{q}_{i})=P\left|\tilde{u}(\boldsymbol{q}_{s}+\boldsymbol{q}_{i})\right|^{2}\label{eq:thin_crystal_far_field_tp}
\end{equation}
\begin{equation}
C^{(1)}(\boldsymbol{q}_{s})=\int d\boldsymbol{\rho}|u(\boldsymbol{\rho})|^{2}=const\label{eq:thin_crystal_far_field_sp}
\end{equation}

Thus the far-field signal coincidence rate for a fixed idler position
position equals the angular spectrum of the pump beam. The resulting
image is shifted by $\boldsymbol{q}_{i}$. In this case the single-photon
rate of both signal and idler is constant and independent of the transverse
position.

Now we consider a finite-length crystal with a plane-wave pump beam.
All other parameters remain unchanged from the previous thin-crystal
model. With both detectors in the far-field plane $p_{3}$, we obtain
the following photon count rates.

\begin{equation}
C^{(2)}(\boldsymbol{q}_{s},\boldsymbol{q}_{i})=\begin{cases}
P|\text{sinc}(\Delta k_{z}(\boldsymbol{q}_{s})L/2)|^{2} & \boldsymbol{q}_{s}=-\boldsymbol{q}_{i}\\
0 & \boldsymbol{q}_{s}\neq-\boldsymbol{q}_{i}
\end{cases}\label{eq:plane_wave_far_field_tp}
\end{equation}
\begin{equation}
C^{(1)}(\boldsymbol{q}_{s})=P|\text{sinc}(\Delta k_{z}(\boldsymbol{q}_{s})L/2)|^{2}\label{eq:plane_wave_far_field_sp}
\end{equation}
Now the coincidence count rate (\ref{eq:plane_wave_far_field_tp})
exhibits perfect anticorrelation with $\boldsymbol{q}_{s}=-\boldsymbol{q}_{i}$
as expected while displaying the typical SPDC ring structure.

Some implicit assumptions were made while deriving Eq. (\ref{eq:two_photon_thin_crystal_near_field})-(\ref{eq:plane_wave_far_field_sp}):
All the formulas given here are only adequate within the limits of
the paraxial approximation. The paraxial approximation is restricted
to small angular frequencies $\boldsymbol{q}$. With the integrals
extending from minus infinity to infinity, the paraxial approximation
is insufficient on a large part of the integration domain. Some of
the integrals involved are divergent. This problem has been addressed
by introducing a cutoff in the integration domain. This cutoff can
be justified due to the limiting apertures that are always present
in an experiment. These factors account for unphysical results such
as the constant single-photon count rate (\ref{eq:thin_crystal_far_field_sp})
at all angles.

\subsection{Numerical Implementation}

\begin{figure}
\begin{centering}
\includegraphics[width=0.7\columnwidth]{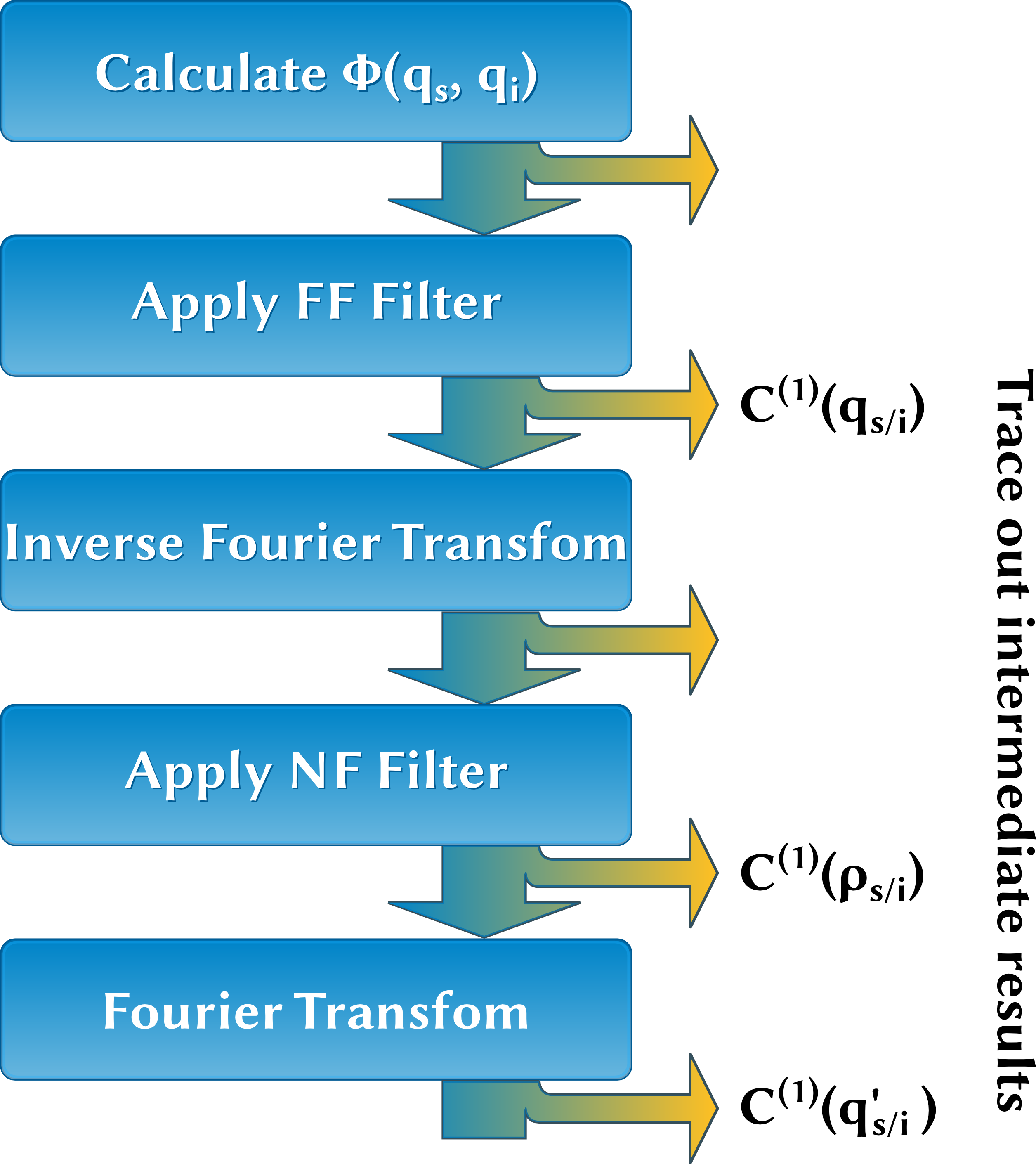}
\par\end{centering}

\caption{Flow diagram of the numerical simulation.}
\end{figure}

For reasons outlined below, we assume without loss of generality that
both the signal and the idler detectors are situated in their respective
far-field planes $p_{3}$ in the measurement stage. As evident from
Eq. (\ref{eq:c2_far_field}), the model can be thought of as a series
of Fourier transforms and filtering operations on a complex four-dimensional
field amplitude that propagates through the experimental setup. At
each stage the intermediate two-dimensional reduced coincidence distributions
$C_{s}^{(1)}$ and $C_{i}^{(1)}$ are stored. Although both detectors
are assumed to be in the far-field, the numerical model yields the
reduced coincidence count distributions for all planes $p_{1}$ to
$p_{3}$ both before and after applying the apertures $\mathcal{T}$,
$\mathcal{N}$. Using appropriate apertures, these reduced coincidence
rates can be used to directly model typical two-photon experiments
as outlined in section \ref{sub:Experimental-Setup-for}.

The principal difficulty of the computation results from the $\mathcal{O}(n^{4})$
memory requirement for holding the (propagated) amplitude $\tilde{\Phi}$
in memory. With the 64 GB of RAM commonly available on today's workstations,
the simulation is thus limited to around 240 points in each of the
four dimensions, assuming all calculations are made in complex double
precision. All transforms are done in-place to conserve memory. Typical
runtimes on a 48-core AMD Opteron System with 64 GB of RAM are around
4 minutes for a single simulation run.

Unfortunately it is impossible to directly partition the amplitude
(\ref{eq:spectral_amplitude}) into tiles and propagate those tiles
independently through the experimental setup. This is due to the ``each
output depends on each input'' property of the Fourier transform.
It is possible however to split up the DFT by using a manual radix-k
decimation in time or frequency. By repeatedly discarding and recalculating
intermediate results, the reduced coincidence count rates can be calculated
at $\mathcal{O}(k^{-4})$ the required memory at the expense of requiring
$\mathcal{O}(k^{4})$ more time for a single DFT. Because the model
requires two DFTs, the time tradeoff for the full propagation is $\mathcal{O}(k^{8})$
while maintaining the $\mathcal{O}(k^{-4})$ memory efficiency. An
additional drawback is the complex bookkeeping required to reassemble
the solution in four dimensions. Using a radix-2 decimation to perform
a simulation with $N=500$ would thus take several days instead of
around one hour while avoiding the need for approximately 940 GB of
RAM.

A viable alternative which we applied is the use of solid state disks
to cache intermediate results that do not fit into core memory. Their
vastly superior access times and number of input/output operations
per second enables efficient caching that is impossible to achieve
with traditional rotational media. By employing two 500 GB SSDs in
a RAID0 configuration, the above-mentioned simulation with $N=500$
takes approximately four days to complete and thus seems to be competitive
to the manual radix-k splitting approach.

Both the Fourier transform and the filtering operation parallelize
trivially and therefore benefit from using multiple threads simultaneously.
The simulation was implemented in C using the FFTW library \citep{frigo_design_2005}
for parallel in-place discrete Fourier transforms and OpenMP for concurrency.
Data postprocessing and visualization was done using matplotlib, numpy
and scipy \citep{hunter_matplotlib:_2007,jones_scipy:_2001,walt_numpy_2011}.

\subsection{Double slit experiment with entangled photons\label{sub:Experimental-Setup-for}}

\begin{figure*}
\begin{centering}
\includegraphics[width=0.9\textwidth]{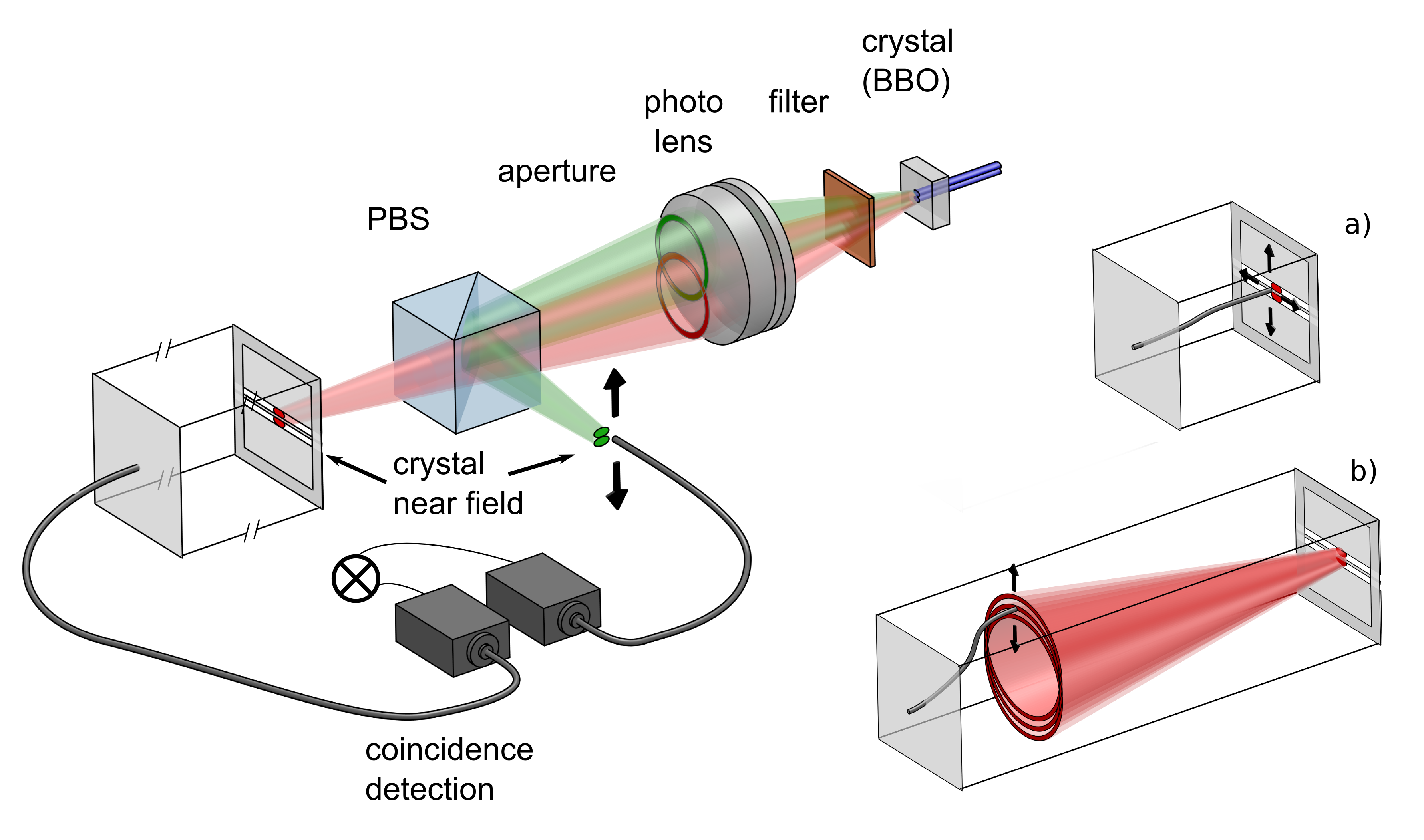}
\par\end{centering}

\caption{\label{fig:schematic-experiment}Schematic view of the experimental
setup. The signal and idler beams are depicted in red and green respectively.
Inset a) demonstrates the position of the signal detector for the
near-field measurement while inset b) shows the corresponding far-field
measurement.}
\end{figure*}

Menzel et. al \citep{menzel_wave-particle_2012} reported a double
slit experiment that measured the interference fringe visibility $V$
and distinguishability $D$ for entangled photons generated by SPDC.
Quantum mechanics mandates that $D^{2}+V^{2}\leq1$, which can be
interpreted as a manifestation of the duality principle \citep{scully_quantum_1991,englert_fringe_1996}.
For brevity we will refer to it as ``DV-inequality'' in the remainder
of the paper. The theoretical model introduced in section \ref{sub:Theoretical-Model}
will be used to quantitatively analyze that experiment. We briefly
review the relevant experimental facts and subsequently describe the
numerical implementation.

The experimental setup is shown in Fig. \ref{fig:schematic_model}.
A 2 mm thick BBO crystal with a phase-matching angle of 42.4 degree
is pumped by a TEM\textsubscript{01} beam at 404 nm. The beam is diffraction-limited
while it's width along the narrow x-axis is 140 $\mu$m. A line-pass
filter at 808 nm with a full-width at half maximum filter width of
2nm is used to block all light except the degenerate signal and idler
photons at 808 nm. The end facet of the crystal is then imaged onto
the near-field plane of the setup using a Nikon AF 50/1.2 photographic
lens. The resulting image is magnified by a factor of 2.2. A polarizing
beam splitter separates the signal and idler beams. A double-slit
with a center-to-center slit separation of 230 $\mu$m and a slit width
of 65 $\mu$m is placed into the near-field plane of the signal path.
The signal detector can be placed directly behind or in the far-field
of the double slit. The idler detector can be positioned everywhere
within the near-field plane. Both detectors use a gradient index multimode
fiber with a core diameter of 62.5 $\mu$m. Alternatively the signal detector
can be swapped out with an Andor Ixon EMCCD single-photon camera.
The camera is used to record the transverse single-photon count distribution.

We perform two different experiments using the described setup. Experiment
one places the Ixon camera in the far-field of the signal beam with
the double slit present at the near-field plane. We then record the
two-dimensional single-photon distribution $C^{(1)}(\boldsymbol{q}_{s}')$
at $p_{3}^{(s)}$ (c.f. Fig. \ref{fig:schematic_model}).

The second experiment is composed of two measurements which we shall
denote 2a and 2b. In experiment 2a, the signal and idler detectors
are both placed in the near-field plane with the double slit in the
signal light path. With the idler detector fixed, the signal detector
is scanned in the x-y plane behind the double slit. We calculate the
which-way distinguishability
\begin{equation}
D=\frac{|C_{upper}^{(1)}-C_{lower}^{(1)}|}{C_{upper}^{(1)}+C_{lower}^{(1)}}
\end{equation}
from the resulting 2-dimensional coincidence distribution. Here $C_{upper}^{(1)}$
is the combined coincidence rate detected at the upper slit
\begin{equation}
C_{upper}^{(1)}=\int\limits _{\text{upper slit}}d\boldsymbol{\rho}_{s}C^{(1)}(\boldsymbol{\rho}_{s})
\end{equation}
with $C_{lower}^{(1)}$ defined analogously. We repeat this experiment
for different idler positions along along the line $\rho_{ix}=0$.
We thus establish a relation between the y-position of the idler detector
and D. After having performed that measurement, we are in a position
to make the following statement: if we detect a coincidence event
with the idler detector at a given position, we can infer with a certain
probability that the associated signal photon has passed through the
lower (or upper) slit. Experiment 2b now places the signal detector
in the far-field of the double slit. We then record a single coincidence
count rate profile along the line $q_{sx}'=0$ and calculate the visibility
$V$ of the interference fringes by fitting the theoretical double
slit interference function
\begin{equation}
C^{(1)}(\xi)=A+B\frac{\text{sinc}^{2}(\xi)}{2}\left[1+V\cos\left(\frac{2a\xi}{b}+\phi\right)\right]
\end{equation}
to the profile \citep{menzel_wave-particle_2012}. The dimensionless
variable $\xi=sq_{sx}'$ depends linearly on the angular momentum
coordinate in the far-field plane $p_{3}^{(s)}$ while the irrelevant
scaling due to the particular implementation of the experiment is
absorbed into the scaling factor $s$. As in 2a, we repeat the experiment
for different positions of the idler detector along $\rho_{ix}=0$.
We therefore obtain the relation between the visibility V of the interference
fringes and the position of the idler detector. Combining the results
from 2a and 2b we effectively measured the relationship between $V$
and $D$ by using the idler position as a proxy measurement of $D$.
Further experimental details can be found in \citep{menzel_wave-particle_2012}.

These experiments can be simulated by placing a set of appropriate
apertures into the numerical model. The numerical parameters for the
double slit and the fiber radius differ from the experimental parameters
to account for the 2.2x magnification factor of the experimental imaging
system which is not present in the numerical simulation. Please refer
to Fig. \ref{fig:schematic_model} and section \ref{sub:Theoretical-Model}
for the naming conventions. The pump beam is a TEM\textsubscript{01}
beam with a $4\sigma$-width of 140 $\mu$m along it's narrow
x-axis. No apertures are required in the far-field plane $p_{1}^{(s/i)}$
of our simulation e.g. $\mathcal{T}_{s/i}\equiv1$. A double slit
with a center-to-center slit separation of 105 $\mu$m and a slit
width of 30 $\mu$m is placed into to signal path at $p_{2}^{(s)}$.

Experiment 1 can be simulated by omitting the idler aperture ($\mathcal{N}_{i}\equiv1$).
Then the resulting numerical $C^{(1)}(\boldsymbol{q}_{s}')$ is equivalent
to the single-photon picture recorded by the Ixon EMCCD. 

Experiments 2a and 2b are emulated by placing a circular aperture
with a radius of $14\mu\text{m}$ in the idler path at $p_{2}^{(i)}$.
The double slit remains in the signal path. Thus the reduced coincidence
count rates $C^{(1)}(\boldsymbol{\rho}_{s})$, $C^{(1)}(\boldsymbol{q}_{s}')$
directly correspond to the respective rates introduced above. The
different idler positions are simulated by displacing the center of
the circular aperture.

\section{Results}

\subsection{Comparison of the numerical model against analytical solution}

\begin{figure}
\begin{centering}
\includegraphics[width=0.5\columnwidth]{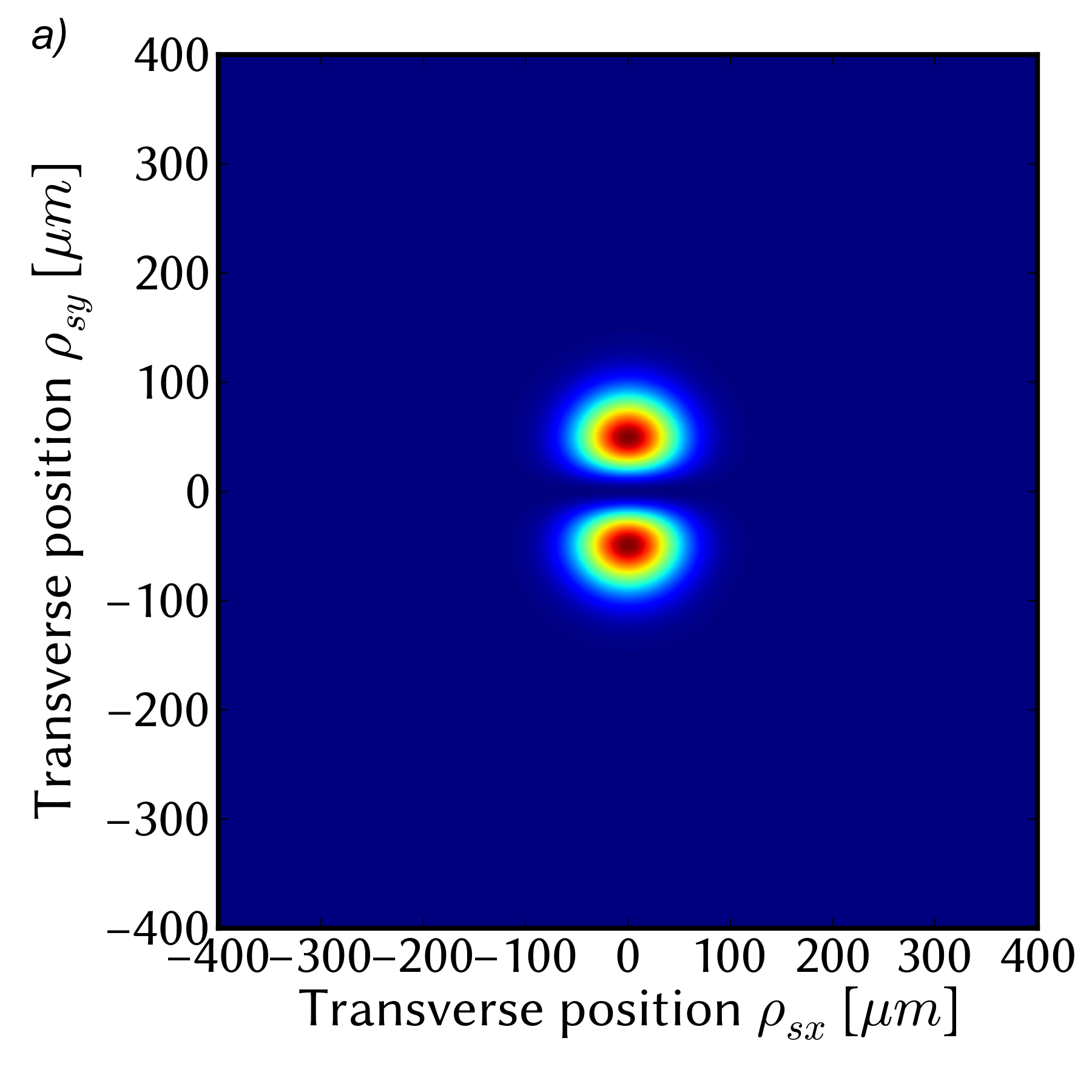}\includegraphics[width=0.5\columnwidth]{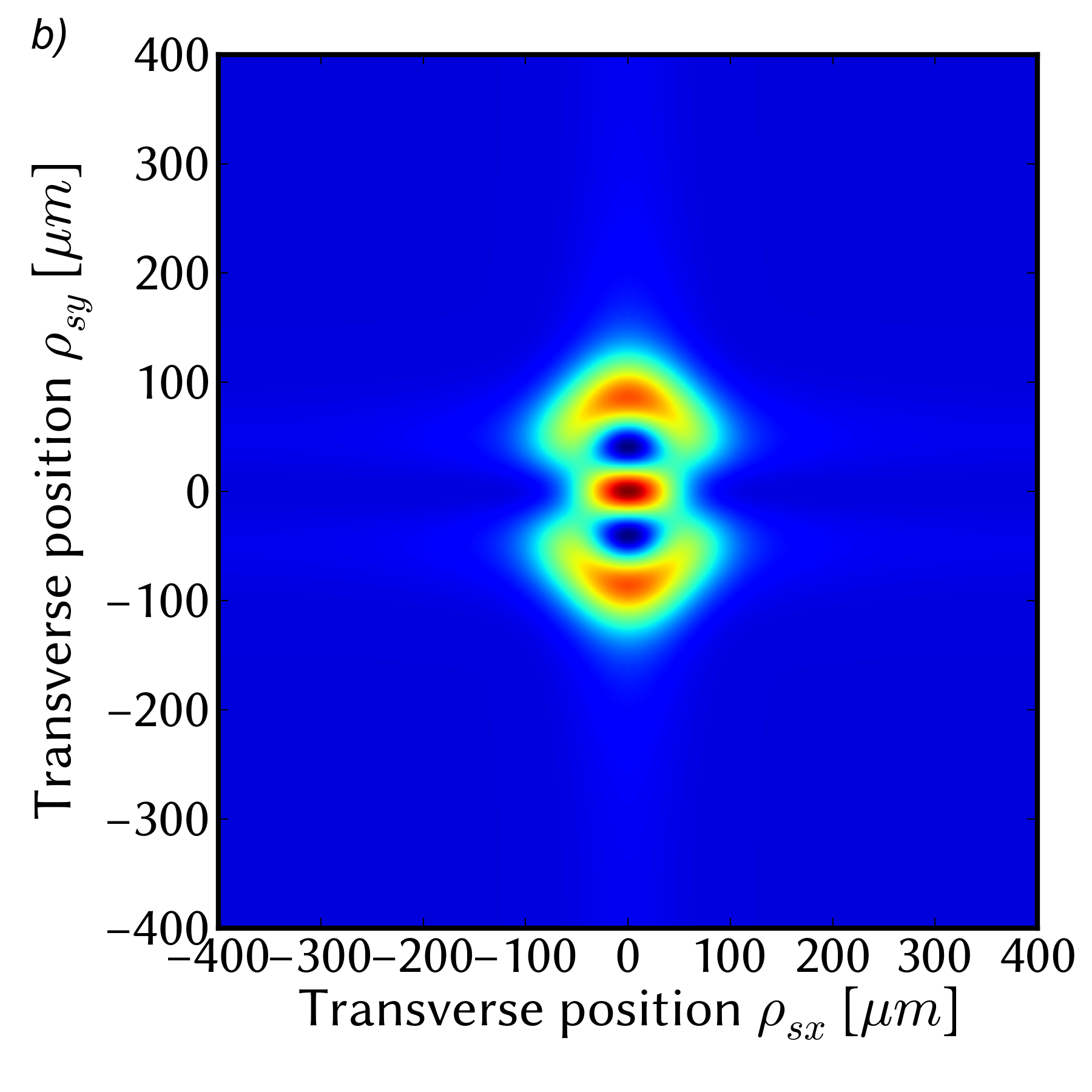}
\par\end{centering}

\caption{a) Analytically calculated signal single-photon distribution $C^{(1)}(\boldsymbol{\rho}_{s})$
in the near-field plane $p_{2}^{(s)}$. b) Deviation between numerical
and analytical result with red indicating a maximum relative error
of approximately $0.005$. \label{fig:comparison_thin_crystal}}

\end{figure}

First we compare the numerical result to the analytical expression
for the coincidence count rate in the thin-crystal approximation (\ref{eq:two_photon_thin_crystal_near_field})-(\ref{eq:thin_crystal_far_field_sp}).
Fig. \ref{fig:comparison_thin_crystal} shows the analytical solution
and the relative error
\begin{equation}
\Delta\epsilon=\frac{\left|C_{analytic}^{(1)}-C_{simulation}^{(1)}\right|}{\max\limits _{\rho}\left(C_{analytic}^{(1)}\right)}
\end{equation}
 of the numerical simulation in the near-field plane. The maximum
relative error is smaller than $5\cdot10^{-3}$ for $N=240$ which
is the maximum resolution for 64 GB RAM. The analytical and numerical
results are virtually indistinguishable. For N=500, which corresponds
to approximately 937 GB of RAM, the maximum relative error decreases
to $4\cdot10^{-3}$. According to Eq. (\ref{eq:thin_crystal_far_field_sp}),
the far-field count rate is expected to be constant everywhere. This
behavior is reproduced by the numerical simulation.

\begin{figure}
\begin{centering}
\includegraphics[width=0.5\columnwidth]{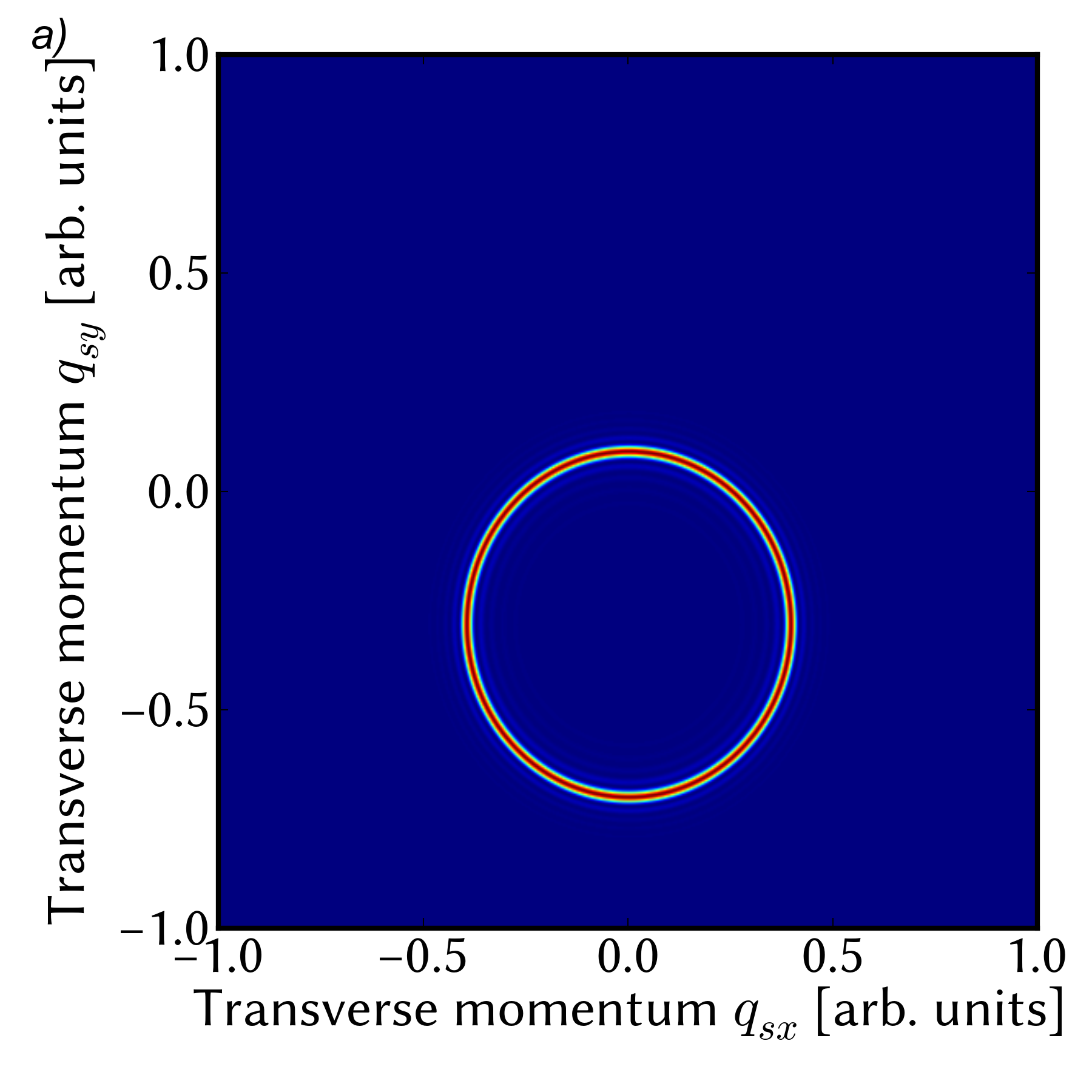}\includegraphics[width=0.5\columnwidth]{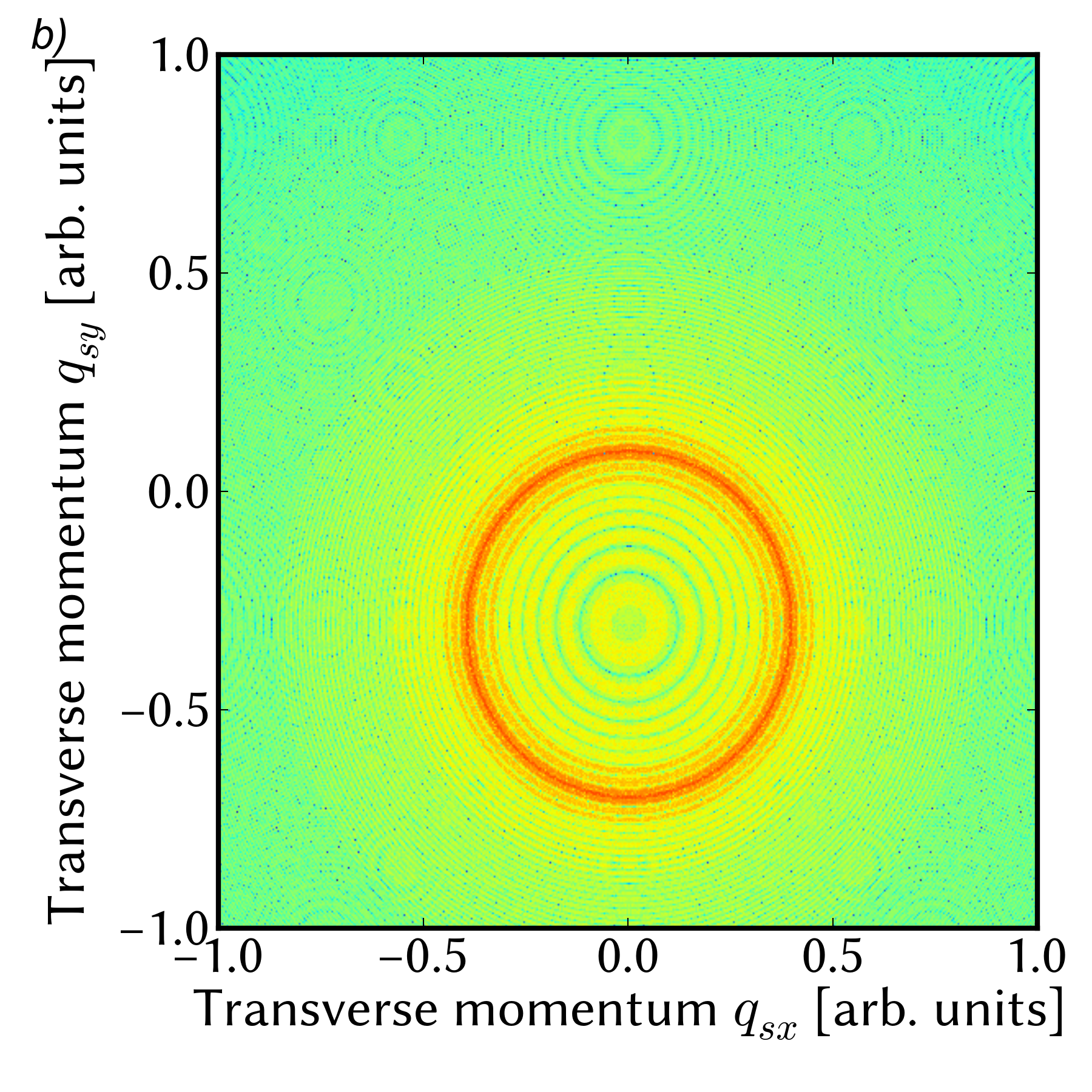}
\par\end{centering}

\caption{a) The signal single-photon distribution in the far-field as calculated
    from the analytic solution for N=500. b) Relative error between
analytic solution and the simulation result on a logarithmic color
scale {[}dB{]}.\label{fig:delta_spectrum}}

\end{figure}

The analytical and numerical results to the plane wave pump field
approximation in the far-field are shown in Fig. \ref{fig:delta_spectrum}.
They exhibit a maximum relative error of the order of $3\cdot10^{-4}$
for both N=240 and N=500.

Furthermore we performed a basic consistency test \citep{boyd_chebyshev_2001}
by gradually increasing the resolution of the numerical simulation.
A consistent simulation must converge towards the true solution as
$N\to\infty$. Thus $\lim{}_{N\to\infty}||\tilde{\Phi}_{N}-\tilde{\Phi}_{N+1}||=0$
is a precondition for consistency. The the relative change with respect
to the simulation's resolution decreases monotonically and reaches
$10^{-4}$ for $N=240$.

\subsection{Experimental results and comparison to the simulation\label{sub:Experimental-results-and}}

\begin{figure}
\begin{centering}
\includegraphics[width=0.5\columnwidth]{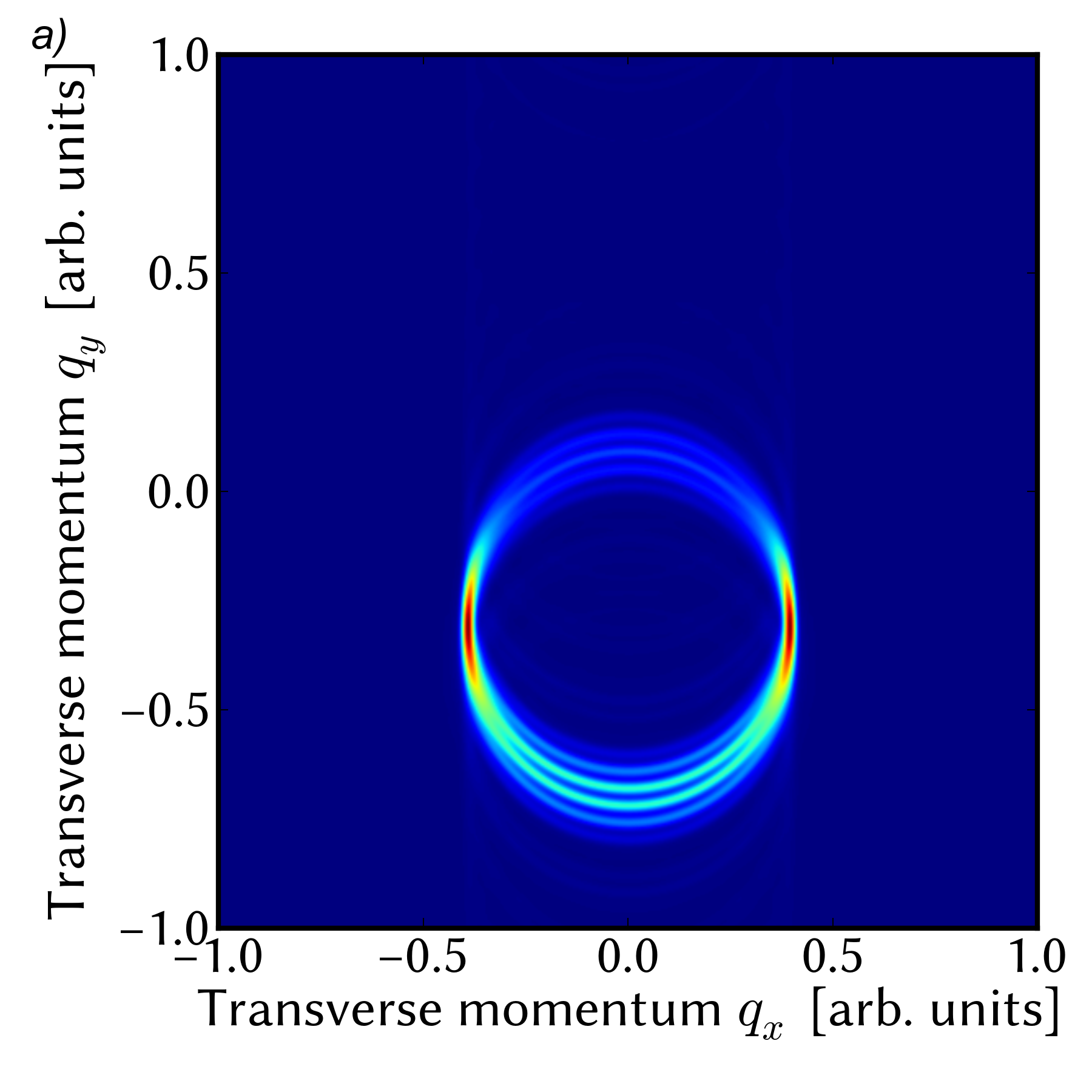}\includegraphics[width=0.5\columnwidth]{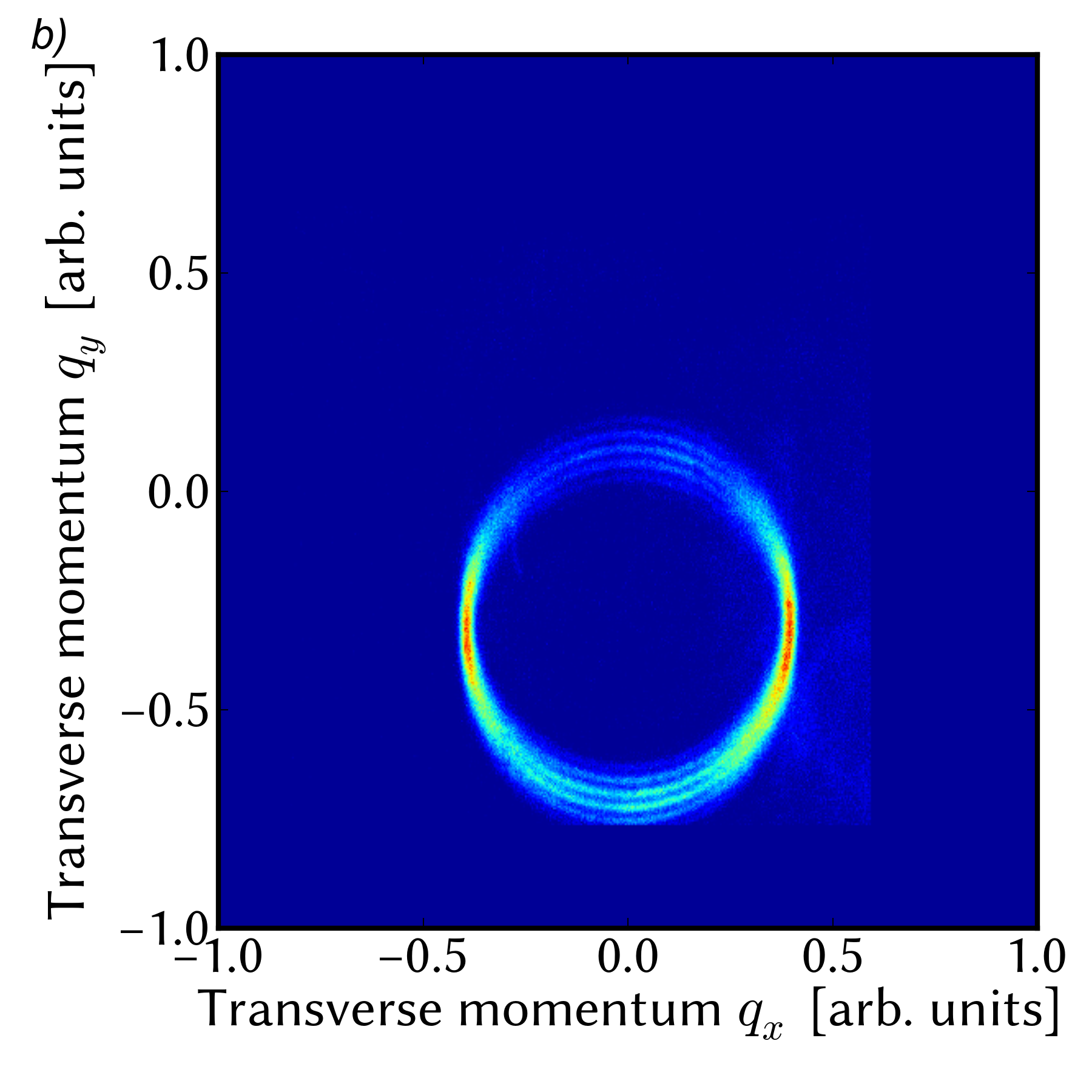}
\par\end{centering}

\caption{a) Single-photon distribution in the far-field of the double slit
    as calculated numerically. b) Measured single-photon distribution
for the corresponding experimental setup.\label{fig:Single-photon-distribution}}

\end{figure}

The experimental single-photon rate for the first experiment outlined
in section \ref{sub:Experimental-Setup-for} is shown in Fig. \ref{fig:Single-photon-distribution}
on the right side. On the left side of the same figure is the result
of the respective simulation. Both the simulation and the experimental
photon count distribution exhibit an uneven number of interference
fringes on the upper side of the ring and an even number on the lower
side. Five interference fringes are clearly visible on the upper side
with six on the lower side. Aside from the pronounced noise floor
in the experimental data, the simulated single-photon count distributions
replicate the experimental data neatly.

\begin{figure}
\begin{centering}
\includegraphics[width=0.5\columnwidth]{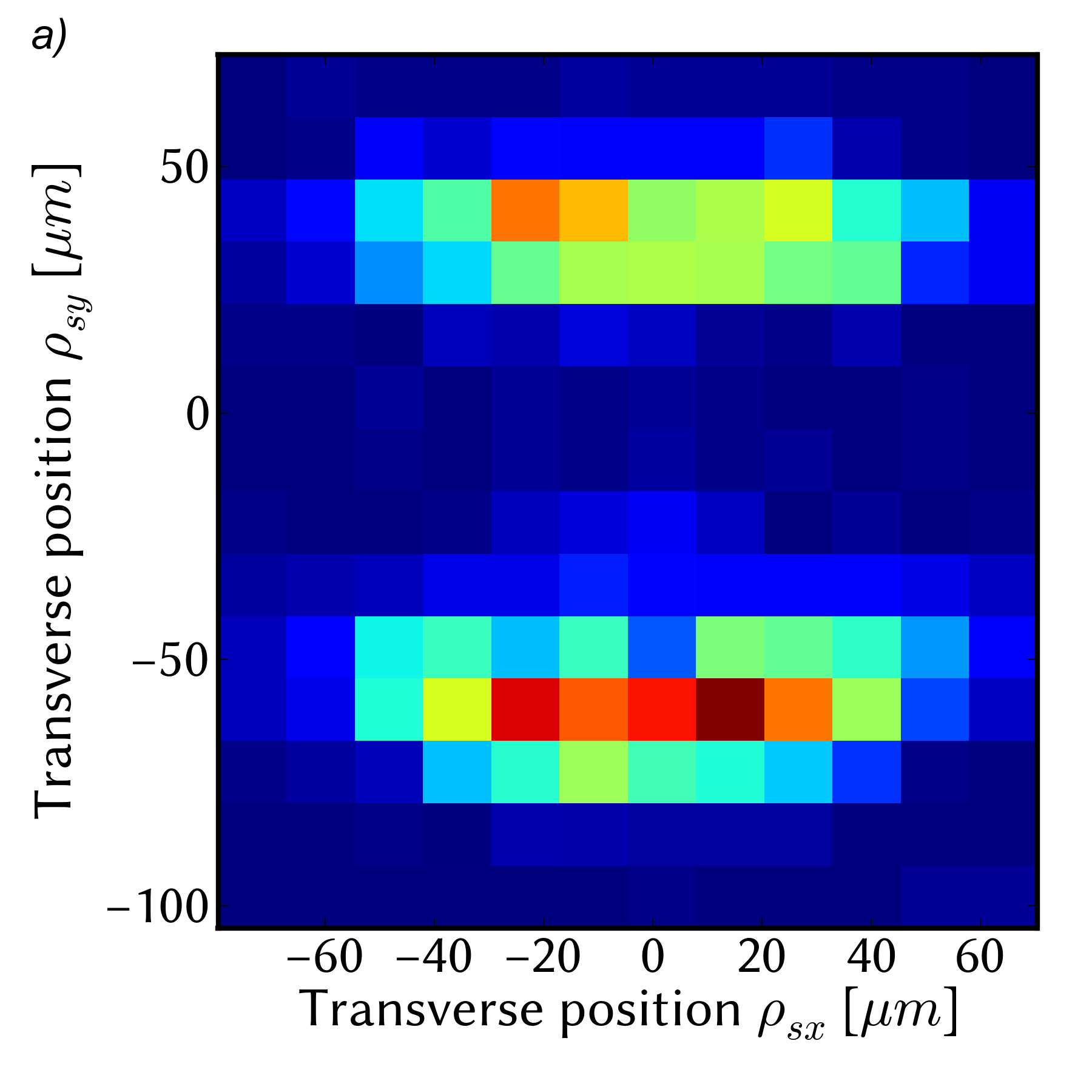}\includegraphics[width=0.5\columnwidth]{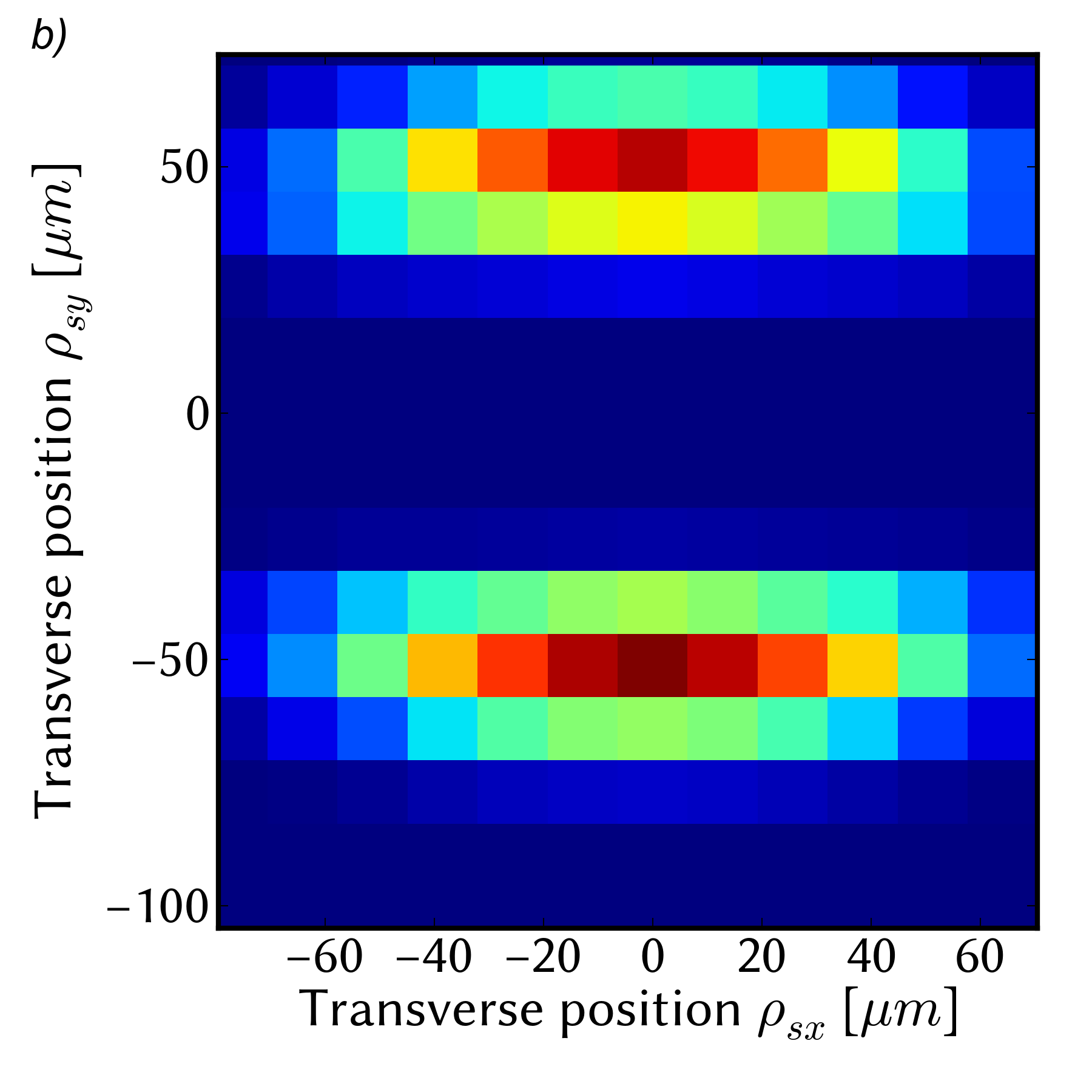}
\par\end{centering}

\begin{centering}
\includegraphics[width=0.5\columnwidth]{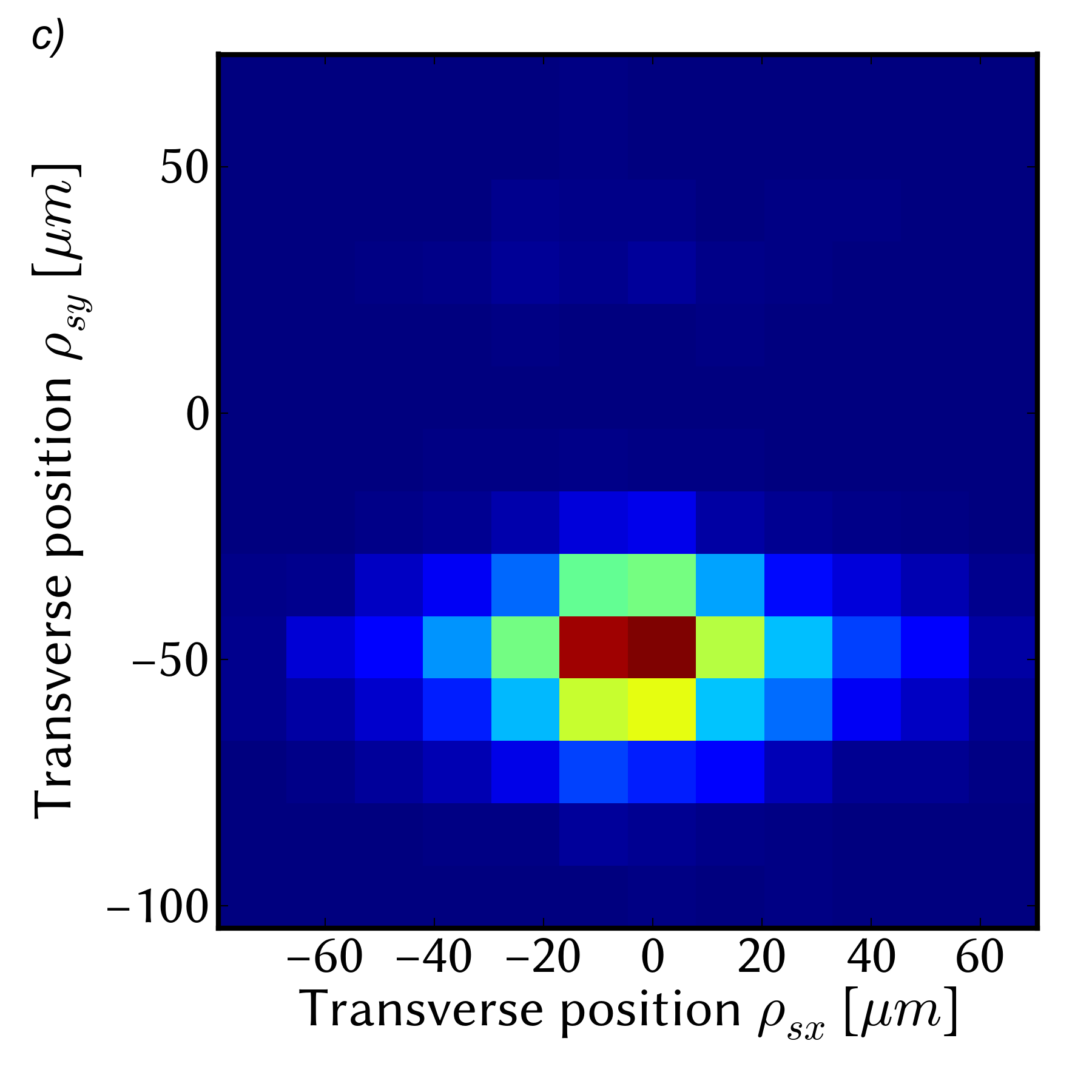}\includegraphics[width=0.5\columnwidth]{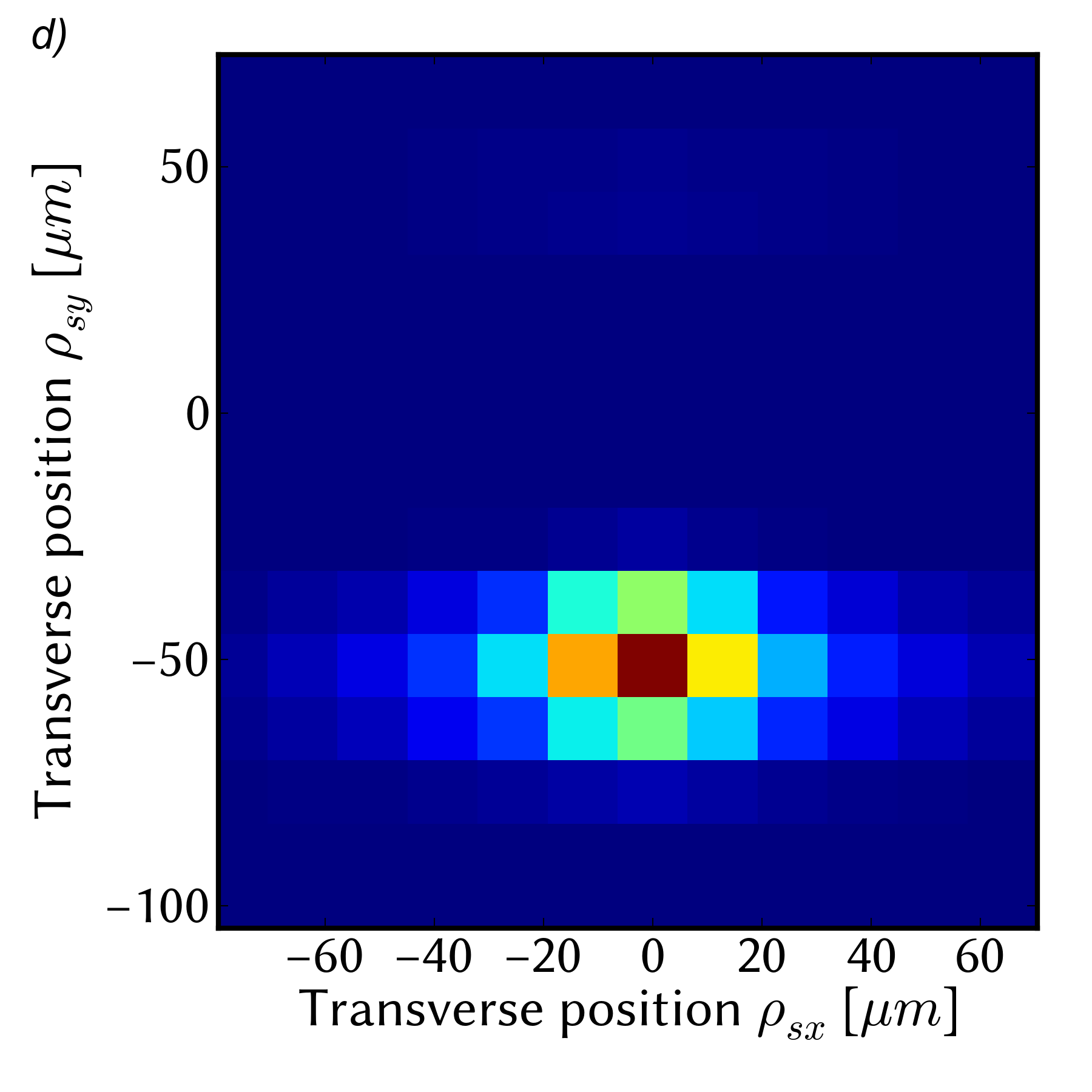}
\par\end{centering}

\begin{centering}
\includegraphics[width=0.5\columnwidth]{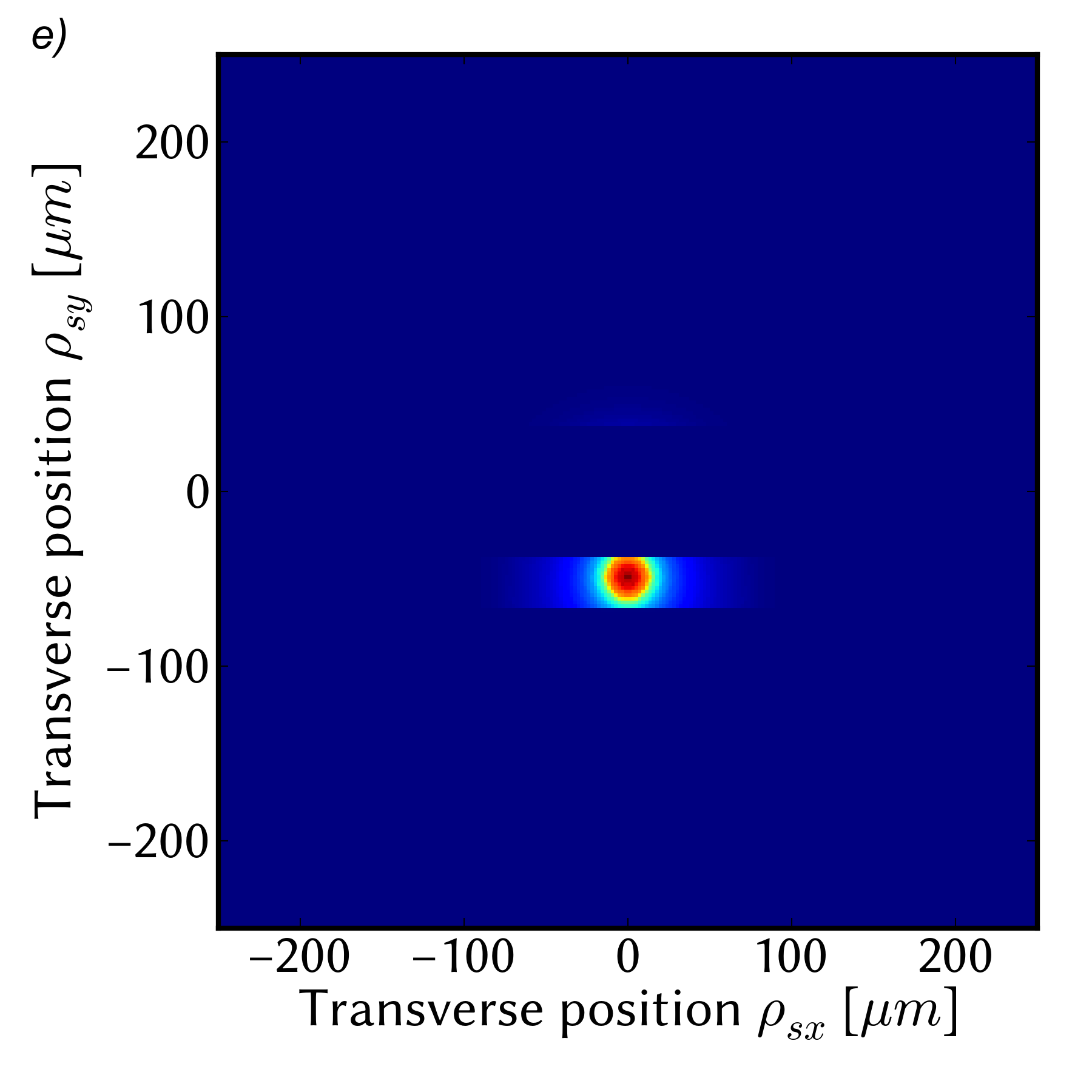}\includegraphics[width=0.5\columnwidth]{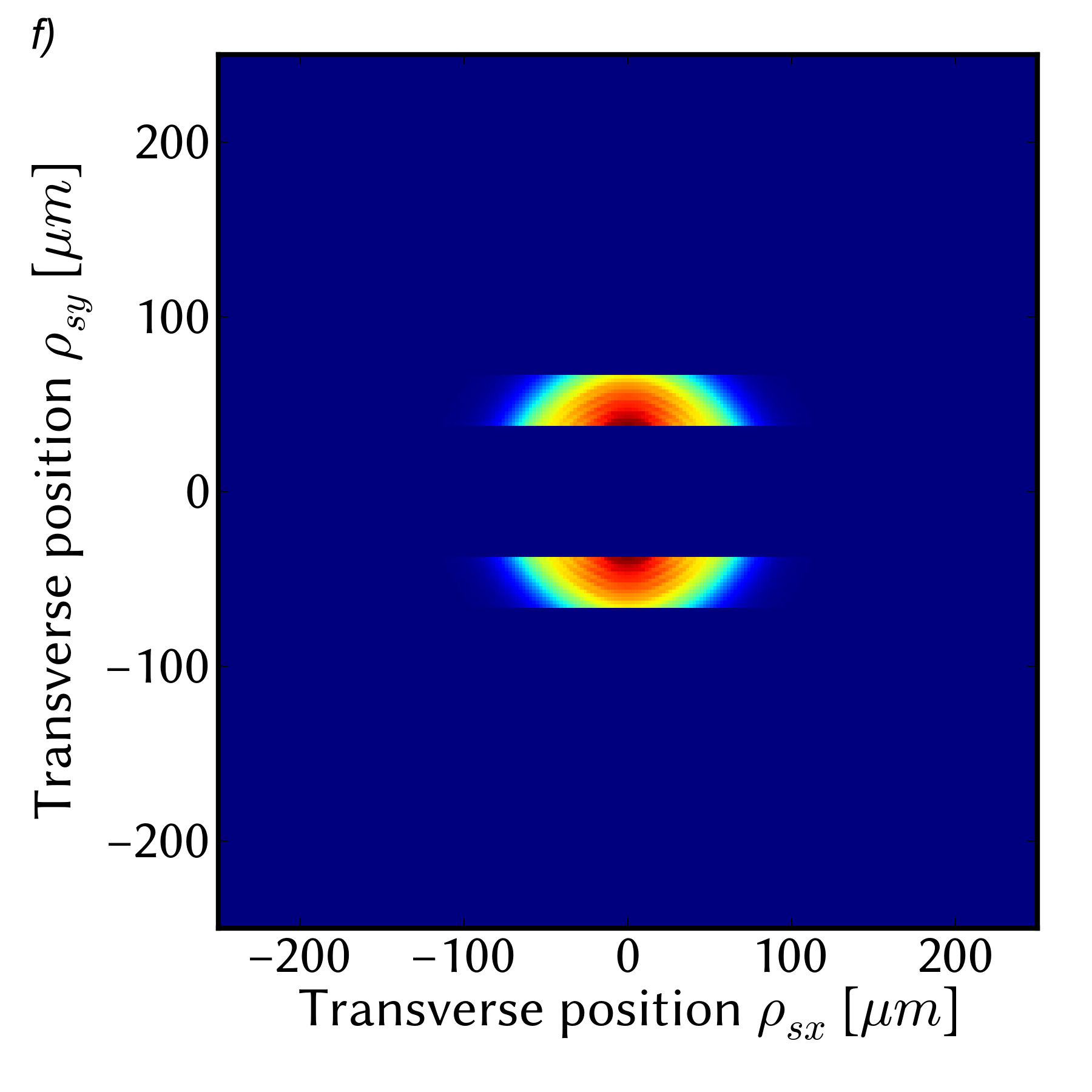}
\par\end{centering}

\caption{a) Near-field signal coincidence distribution behind the double slit
with the idler detector placed in the middle between the two lobes.
b) Simulated distribution with equivalent parameters to a). c) Coincidence
distribution with the idler detector placed at the lower lobe of the
TEM\textsubscript{01} pump beam. d) Simulated distribution with equivalent
parameters to c). e) Full-resolution simulation result for d). f)
Full-resolution simulation result for b). The count rates are given
in arbitrary units. Each plot is normalized to unity.\label{fig:a)-Near-field-signal}}

\end{figure}

We now present selected results for the experiments 2a and 2b. For
the idler detector placed in between the two lobes of the TEM\textsubscript{01}
mode at $(\rho_{ix},\rho_{iy})=(0,0)$, the near field coincidence
distribution for the signal detector together with the numerical results
can be seen in Fig \ref{fig:a)-Near-field-signal}.a and \ref{fig:a)-Near-field-signal}.b.
This idler position results in a distinguishability of $D=0.0$. In
addition to the coordinate transform mentioned above, the numerical
result was convolved with a circular aperture kernel with a radius
of 14 $\mu$m to emulate the signal fiber response. The result
of the convolution was then downsampled to match the experimental
sample spacing. The numerical simulation result without any transformations
is shown in Fig. \ref{fig:a)-Near-field-signal}.f. Note that the
finite signal fiber size must be included explicitly in a post-processing
step while the idler fiber size is implicitly included in $C^{(1)}(\boldsymbol{\rho}_{s})$.
Without this postprocessing, the area of the signal detector equals
the intrinsic resolution limit of the simulation which in this case
is $\Delta\rho_{sx}=\Delta\rho_{sy}\approx2.1\ \mu\text{m}$.

If the idler detector is instead placed at $\rho_{iy}=-48\ \mu\text{m}$
which roughly corresponds to the middle of the lower slit, the distinguishability
increases to $D\approx0.96$ for both the simulated and experimental
data. The experimental and numerical coincidence distributions are
shown in Fig. \ref{fig:a)-Near-field-signal}.c and Fig. \ref{fig:a)-Near-field-signal}.d
respectively with the non-transformed numerical result in Fig. \ref{fig:a)-Near-field-signal}.

\begin{figure}
\begin{centering}
\includegraphics[width=0.5\columnwidth]{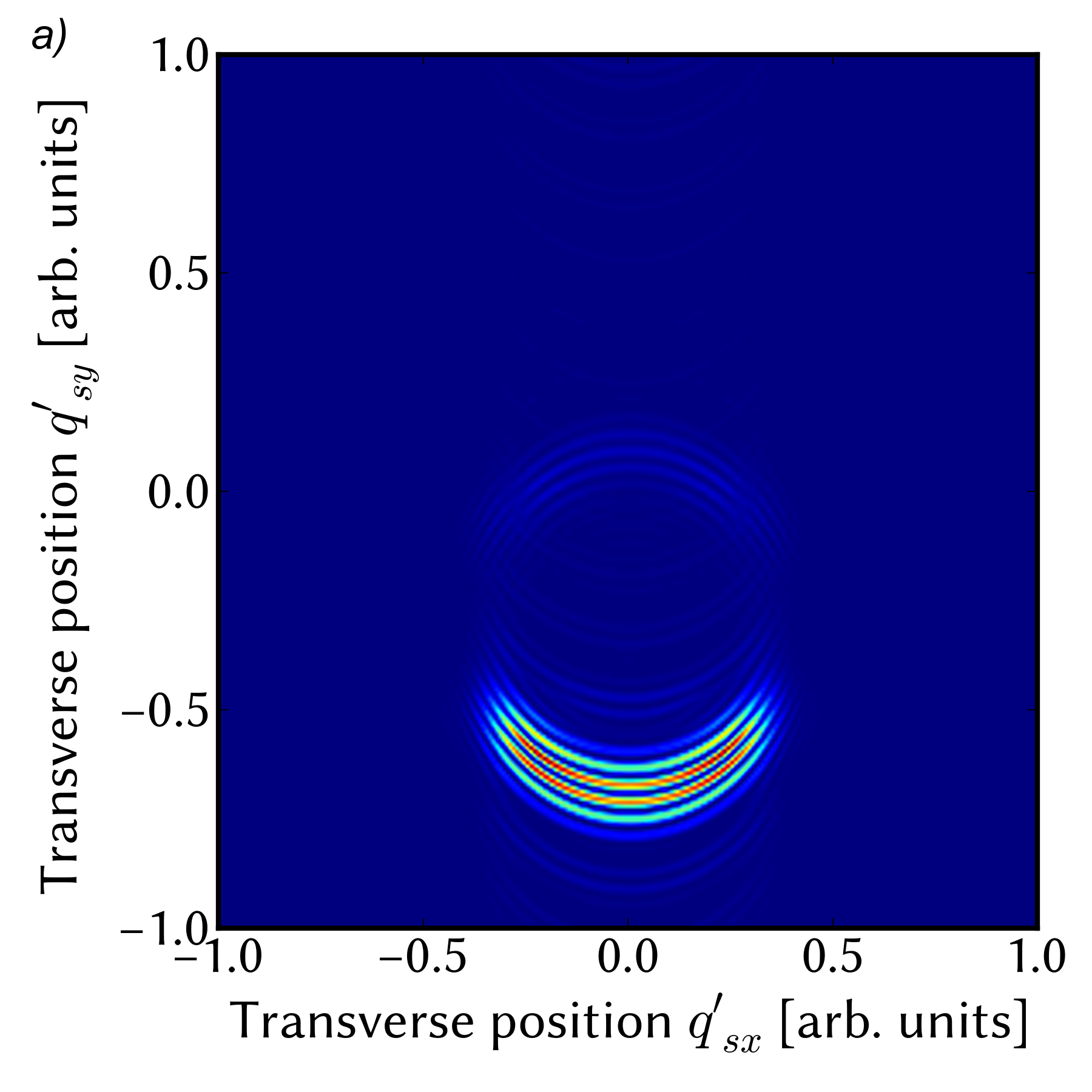}\includegraphics[width=0.5\columnwidth]{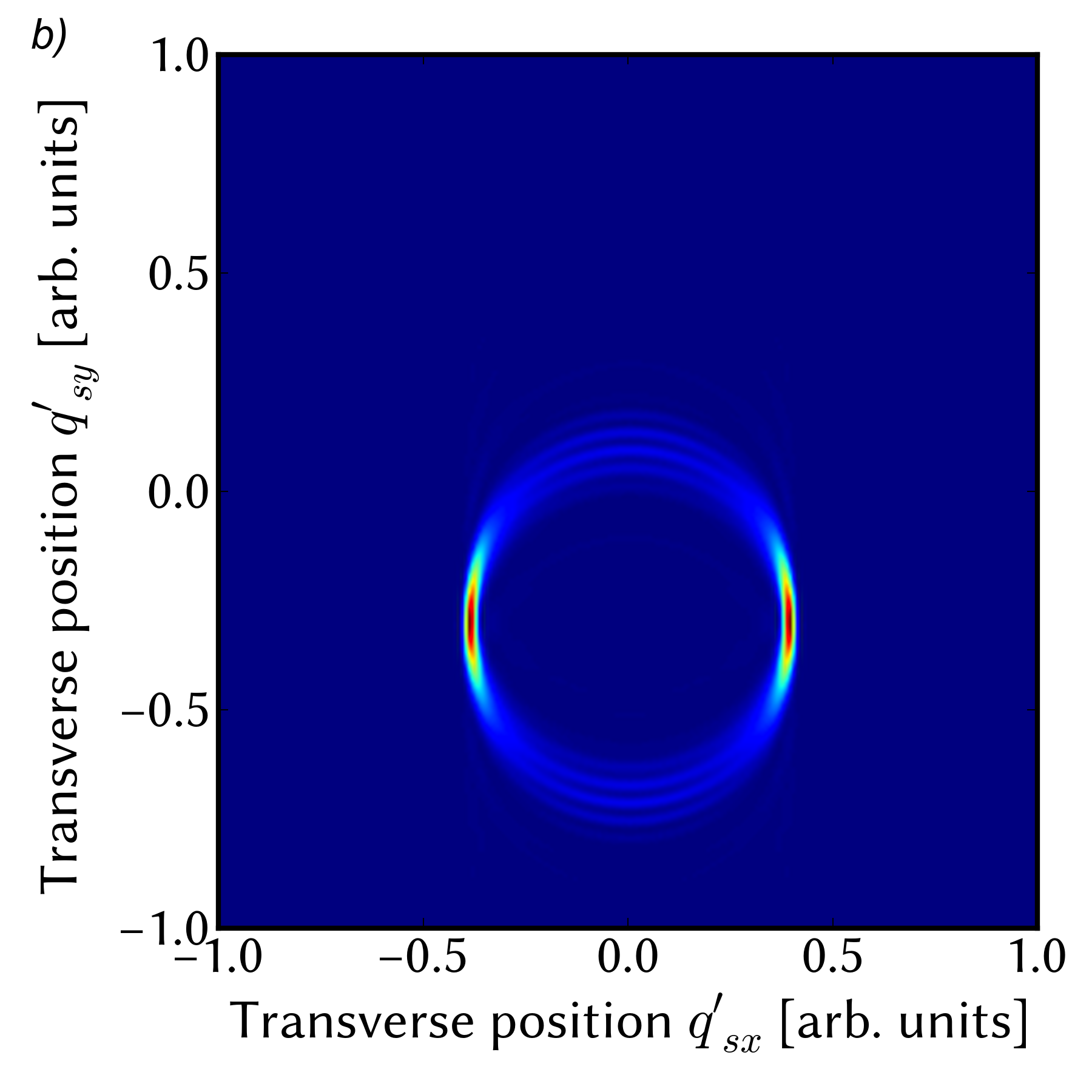}
\par\end{centering}

\caption{a) Far-field coincidence interference pattern in the signal beam
    with the idler detector placed between the two lobes. b) Far-field
interference pattern with the idler detector placed approximately
at the center of the lower lobe. Red indicates a high count rate.\label{fig:Left:-far-field-coincidence}}

\end{figure}

Moving the signal detector into the far field and the idler detector
back to $\rho_{iy}=0$ we get the simulated coincidence distribution
which is displayed in Fig. \ref{fig:Left:-far-field-coincidence}
on the left. The simulated and measured interference fringes along
the line $q_{sx}=0$ through the lower part of the ring result in
visibilities of $V_{sim}=1.0$ and $V_{exp}=0.85$ respectively. This
idler detector position corresponds to the figures \ref{fig:a)-Near-field-signal}
a) and b). If the idler detector is now placed again at $\rho_{iy}=-48\ \mu\text{m}$,
the visibility of the interference fringes drop to $V_{exp}=0.48$
and $V_{sim}=0.58$. The simulated coincidence distribution is shown
in Fig. \ref{fig:Left:-far-field-coincidence} on the right. The color
scales of the two plots in Fig. are individually normalized to unity.
The maximum and overall coincidence count rates on the left plot are
much lower when compared to the right plot. A common color scale would
make the left plot indiscernible. The visibility of the interference
fringes is unaffected by these absolute differences in the coincidence
count rates in the absence of an additive noise floor.

The combined results from both the experiments 2a and 2b are shown
in Fig. \ref{fig:Knowledge-K,-visibility}. The upper subfigure shows
the simulated distinguishability and visibility while the measured
data can be seen in the lower plot. The simulation predicts that the
term $D^{2}+V^{2}$ takes on it's maximum value of $1.47$ at $\rho_{iy}=-35\ \mu\text{m}$.
Experimentally we measured $D^{2}+V^{2}=1.41$ at $\rho_{ix}=-36.5\ \mu\text{m}$. 

\begin{figure}
\begin{centering}
\includegraphics[width=1\columnwidth]{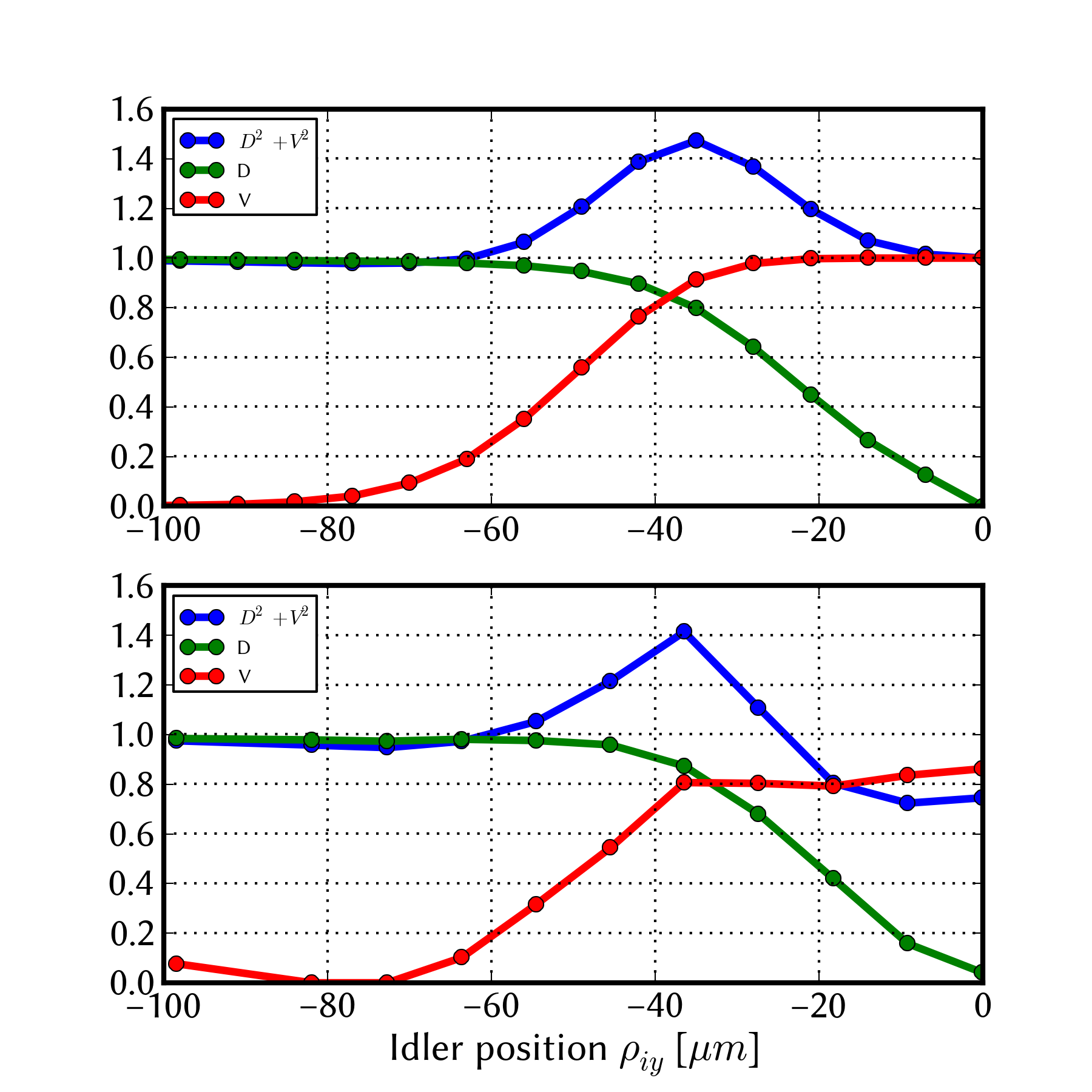}
\par\end{centering}

\caption{Distinguishability $D$, visibility $V$ and $D^{2}+V^{2}$ plotted
over the position of the idler detector. Top: numerical results. Bottom:
experimental data.\label{fig:Knowledge-K,-visibility}}
\end{figure}

\section{Discussion}

The numerical calculations are in agreement with the analytical solutions.
The analytical solutions are a test to confirm if the numerical model
is able to handle simple scenarios. While the simplified analytical
solutions are interesting in their own right, they have several shortcomings
that make them ill-suited to predict our experimental data. The thin-crystal
approximation suffers from a total absence of the phase-matching properties
of the nonlinear crystal. This in turn leads to a far-field pattern
that bears no resemblance to the experimentally observed count distributions,
thus preventing the simulation of interference fringes that were observed
in experiments one and two. The plane-wave pump field approximation
suffers from a uniform near-field distribution that neglects the features
of the pump beam. Our more detailed numerical model does not suffer
from those limitations. Furthermore the numerical model is easily
adaptable to different experimental setups by inserting the appropriate
aperture functions $\mathcal{T}$ and $\mathcal{N}$ without the need
to reevaluate the aperture-dependent convolutions (\ref{eq:c2_far_field}).
We found a resolution of $N=240$ to be sufficiently accurate to make
meaningful predictions about the experimental outcome. The resolutions
up to $N=500$ available with SSD caching might be required for large
pump beams to retain sufficient range and resolution in both the spatial
and angular momentum domains.

We simulated an optimized rerun of a recent experiment by Menzel at
al. to test our numerical simulation and found the model to be in
agreement with experimental data. The far-field single-photon count
distribution from Fig. \ref{fig:Single-photon-distribution} closely
matches the simulated distribution. Minor discrepancies can be attributed
to the experimental noise floor. Menzel et al. \citep{menzel_wave-particle_2012}
previously reported a maximum value of $D^{2}+V^{2}=1.3$. Our new
experiment improved on these results with $D^{2}+V^{2}=1.41$. This
apparent violation of the DV-inequality was predicted by the numerical
simulation which established an upper bound of $D^{2}+V^{2}=1.47$.
Thus the experimental result is close to the predicted theoretical
maximum for the given experimental parameters. The limiting factor
hereby is the insufficient experimental resolution in the far-field
measurement of the interference fringe visibility due to the signal
detector fiber size.

Upon publication of \citep{menzel_two-photon_2013,menzel_wave-particle_2012}
it has been suggested that the experimental results were due to ``leaking''
photons that were not detected in the near-field measurement. In light
of the experimental and theoretical results presented in this paper,
we feel confident to reject the prospect of experimental errors in
our experiment and \citep{menzel_two-photon_2013,menzel_wave-particle_2012}.
While there are ``leaking'' photons, they alone are insufficient
to explain the observed visibilities. 

Note that we make no claim that the DV-inequality is invalid or that
our experiment actually circumvents the duality principle. We merely
demonstrated that the measurements for the given experimental setup
agree excellently with theoretical predictions. A possible explanation
for this mildly surprising results based on the principle of fair
sampling was outlined in \citep{leach_duality_2014}. Our numerical
model provides a complete two-dimensional description of the double-slit
experiment which we will use to further investigate the implications
of fair sampling.

\section{Conclusion}

We developed and implemented a quantitative model to describe the
transverse spatial structure of the light fields generated by SPDC
in the near and far field cases in detail. The model includes the
effects of the crystal parameters such as length and phase matching
angle and handles arbitrary input beams. Arbitrary amplitude manipulations
of the far- and near-fields are possible in a separate preparation
stage.

We repeated a double-slit experiment by Menzel et al. and improved
on these results by measuring an interference fringe visibility of
$V=0.8$ with a simultaneous distinguishability at the double-slit
of $D=0.87$, resulting in an apparent violation of the DV-inequality
with $D^{2}+V^{2}=1.41$. This is close to our predicted theoretical
maximum of $D^{2}+V^{2}=1.47$. We used this experiment to test our
numerical model against experimental data. We found the model to be
in excellent agreement with the measured data.

Thus the numerical model can now be used to test new ideas and identify
interesting parameter ranges before actually committing time to experimental
work. Furthermore the numerical data can be a valuable tool to interpret
the sometimes surprising results arising from experiments with entangled
photons. It will provide detailed insights into the fair sampling
problem at a double slit which will be discussed in a forthcoming
paper.
\begin{acknowledgments}
Robert Elsner is grateful to Rainer Herbst for providing technical
support related to the compute cluster at the University of Potsdam.
\end{acknowledgments}

\bibliography{methodenpaper}

\end{document}